\documentclass[showpacs,aps]{revtex4}

\usepackage{graphicx}
\usepackage{psfrag}

\begin{document}
\bibliographystyle{/home/anogga/.TeX/apsrev}

\title{The $\alpha$ particle based on modern nuclear forces}
\author{A.~Nogga$^{1}$, H.~Kamada$^{2}$, W.~Gl\"ockle$^{3}$, B.R.~Barrett$^{1}$}
\affiliation{
$^{1}$Department of Physics, University of Arizona, Tucson,Arizona 85721, USA
\email{anogga@physics.arizona.edu} \\
$^2$Department of Physics, Faculty of Engineering, Kyushu Institute of Technology, Kitakyushu 804-8550, Japan \\
$^3$Institut f\"ur theoretische Physik II, Ruhr-Universit\"at Bochum,
D-44780 Bochum, Germany}

\date{\today}

\begin{abstract} 
The Faddeev-Yakubovsky equations for the $\alpha$ particle are solved.
Accurate results are obtained for several modern 
NN interaction models, which include charge-symmetry breaking 
effects in the NN force, nucleon mass dependences as well as the Coulomb interaction. 
These models are augmented by three-nucleon forces of different types
and adjusted to the 3N binding energy. Our results are 
close to the experimental binding energy with a slight 
overbinding. Thus there is only little room left for the 
contribution of possible 4N interactions to the $\alpha$ particle
binding energy. We also discuss model 
dependences of the binding energies and the wave functions. 
\end{abstract}


\maketitle

\section{Introduction}

In spite of the tremendously increased computational power of today's
super computers, numerical investigations of nuclear bound states are
still a challenging problem, even for systems of few nucleons. 
Investigations promise insights into the rich structure of nuclear
interactions. To this aim one requires reliable solutions of the
dynamical equations. In this article we would like to present 
results for the $\alpha$ particle, which are based on realistic
microscopic nuclear forces including three-body interactions.

In recent years forces could be adjusted accurately to the huge
amount of available NN low energy scattering data
\cite{cdbonn,av18,nijm93}. The overall agreement of the predictions of
these model forces with the data is essentially perfect. As a result it has
been shown that most of the observables in the low energy regime of
the 3N continuum could be predicted model independently
\cite{report,witala01a}, though the interactions themselves are quite
different. On the other side, it is known since quite a long time 
that the 3N binding energies (BE) are quite model dependent and, moreover, 
are generally smaller than the experimental value 
\cite{friar88b,wu93,stadler91,nogga97}. It is assumed that most of this
underbinding is due to three-nucleon forces (3NF) and modifications 
of the NN interaction in the 
presence of a third nucleon. The latter one is of course also part 
of a three-nucleon force mechanism.

The nature of these 3NF's is still not completely understood. It is clear that
such forces should already arise because of the composite structure of the
nucleons, what is partially taken into account by allowing for intermediate
$\Delta$-excitation. Other mechanisms of various meson-exchange types 
will also contribute (for a review see \cite{robilotta87}).
In recent years there has been new progress in understanding the form
of nuclear forces, because of the application of chiral
perturbation theory ($\chi$PT) 
\cite{weinberg90,weinberg91,kolck94,ordonez96,kaiser97,epelbaum00,epelbaumphd}. 
From this developments 
one can expect a more systematic  understanding of the form of NN and 3N 
forces in the near future. However, $\chi$PT implies 
{\it a priori} unknown constants, the low-energy constants, which have to be
determined from experimental data. The bound states of few-nucleons 
seem to be an ideal laboratory to determine 3NF parameters, as the 
BE's are sensitive to the 3N interaction and they 
are expected to be governed by the low-energy regime of nuclear 
physics \cite{bedaque00}. Therefore, the understanding of nuclear 
bound states is an important contribution to the understanding of 
the 3NF. 

At present these chiral interactions are not as accurate as the
traditional, phenomenological NN forces \cite{cdbonn,av18,nijm93}.
At the order of the chiral expansion parameter
considered up to now \cite{epelbaum01c}, they do not yet describe 
the NN phase shifts with the same accuracy.
Allowing, however, for additional fine tuning a high accuracy 
description can be achieved \cite{entem01}. 
The aim of this
article is to pin down model dependences 
of predictions for the $\alpha$ particle BE and wave function (WF)
properties. To insure that differences in the predictions are not 
due to an inaccurate description of the NN system, but are due to the
more fundamental differences of the interaction models, we restrict
ourselves to the traditional models in this paper. The techniques
developed, however, will help to apply also  the upcoming chiral
interactions. First investigations, using chiral interactions,
have already been undertaken \cite{epelbaum01a,epelbaum01c}.

Our approach leads immediately to a basic problem. 
It has been shown that the 3N interaction cannot be determined
uniquely and, moreover, that each NN interaction has to be accompanied 
by a different 3NF \cite{polyzou90}. For the traditional NN 
interactions, there are no 3NF's available, which have been derived 
consistently to them. Therefore, we have to rely 
on 3NF's, which just take parts of the mechanisms into account, 
which are expected to contribute to the 3NF. These models are, for
example, the Tucson-Melbourne (TM) \cite{coon79} 
and the Urbana IX (Urb-IX) \cite{pudliner97} 3NF's. For the different 
NN interactions these models have been adjusted separately 
to the experimental $^3$H BE, as described in Section~\ref{sec:resa}.
This scheme is justified for two reasons:
\begin{itemize} 
\item It has been shown that many 3N scattering 
      observables in the low energy regime 
 (below $\approx 10$~MeV nucleon lab energy) scale with the $^3$H
BE. This means that predictions for different model Hamiltonians are
equal, when the models predict the same $^3$H 
BE \cite{huber98b,witala98c,friar86b}. An adjustment of the 3NF's 
exclude model dependences related to this phenomenon. We will see that these 
effects are also visible for the $\alpha$ particle. 
\item In the high energy regime (above $\approx 100$~MeV nucleon lab
energy) the predictions for 3N scattering observables are sensitive
to the 3NF showing that the available 3NF models 
are quite different \cite{witala01a}. This insures that the 
models applied in this paper cover a wide range of possible 3NF's. 
Therefore, our results show the model dependences of our current
understanding of the $\alpha$ particle, which are related to the structure 
of the 3NF.         
\end{itemize}

Bound states of light nuclei have been investigated by several groups 
using different techniques 
\cite{wiringa00,pieper01,navratil00,suzukilec,viviani92,
kameyama89,barnea00,ciesielski98,kamada92a,nogga00,kamada01c}.
But much of the work is still 
restricted to somewhat simplified interactions. Perhaps the most advanced 
calculations covering several nuclei 
have been performed by the Argonne-Los~Alamos collaboration
\cite{wiringa00,pieper01}. Using the Greens functions Monte Carlo (GFMC) 
technique, they were able to predict BE's for the light nuclei 
up to $A=8$. However, 
their work is restricted to the AV18 NN interaction model 
and the class of Urb-IX 3NF's 
(new terms not considered here have been added 
in \cite{pieper01}). 
This leads to the question, whether 
the other available interactions give similar or different results 
for these nuclei. In this respect the ``no-core'' shell model 
approach (NCSM) \cite{navratil00} might be more flexible. 
But the work on 3NF's has not been finished 
yet. Therefore, we think that a study of the 4N system can provide 
important new information on the nuclear interactions, if one can investigate 
a wide range of NN and 3N models in this system. 

In this paper we 
use the Faddeev-Yakubovsky scheme to solve the non-relativistic 
Schr\"odinger equation (SE) for four nucleons. This has been started already in 
Refs.~\cite{kamada92a,glockle93a,glockle93b,kamada92b,kamada94b,kamada94c}.
With this method we are able 
to get reliable results for the BE and the WF of the 
$\alpha$ particle for several NN and 3N interactions. The calculations are restricted to $A=3$ and $A=4$, but we were able to pin down the dependence 
on today's interaction models. 
The highly accurate WF, which result from the calculations,
are necessary for the analysis of several on-going or planned experiments on 
the $\alpha$ particle, which might reveal the short-range correlations 
in nuclei \cite{e97-111} or give insights into the charge independence 
breaking of the nuclear interaction \cite{kolckpriv}. Exact WF's 
are also necessary to understand the results of parity violating 
e$^-$ scattering experiments \cite{ramavataram94}. 
Therefore, we will also give first  
results of calculations including the isospin $T=1$ and $T=2$ component 
of the $\alpha$ particle ground state WF.  

Another important issue is a first estimate of the size of 
a possible 4N interaction. We expect that 
it should show up especially prominent in the $\alpha$ particle, because 
of its high density. Our calculations give some hints, as to whether there is 
room for an important contribution of the 4N interaction in nuclei
given today's NN and 3N interaction models.

In Section~\ref{sec:form} we
briefly review the 4N Faddeev-Yakubovsky formalism.
The calculations are based on adjusted 3NF's. The adjustment 
procedure is described in Section~\ref{sec:resa}. 
Our results for the BE's of the $\alpha$
particle based on   various nuclear force combinations are given in 
Section~\ref{sec:resb} and the properties of the obtained 
WF's are presented in Section~\ref{sec:resc}. 
Finally we summarize in Section~\ref{sec:concl}.

\section{The 4N Yakubovsky formalism}
\label{sec:form}

\begin{figure}

\begin{center}
\includegraphics[angle=0,width=7cm]{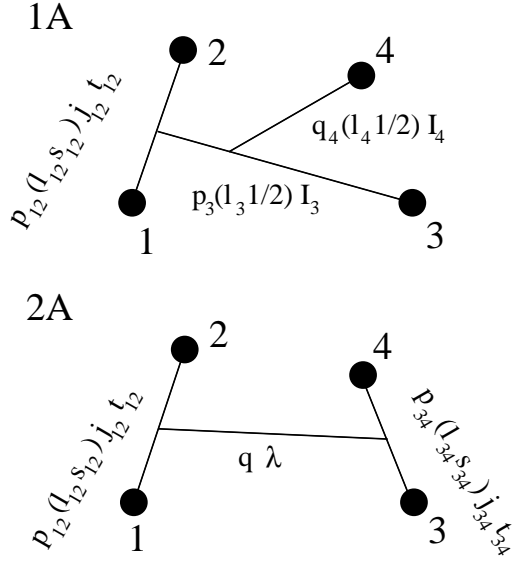}	
\end{center}

\caption{Definition of the 1A and 2A type of Jacobi coordinates.}

\label{fig:jacobi}

\end{figure}

\begin{figure}

\begin{center}

\includegraphics[angle=0,width=7cm]{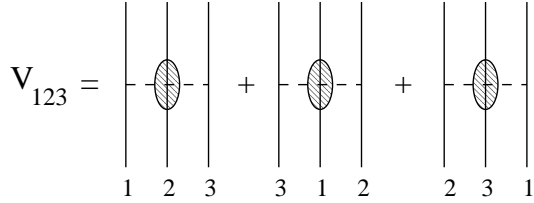}	

\end{center}

\caption{The three parts of a meson exchange 3NF, which differ only by an exchange 
         of the particles.}
\label{fig:3nf}

\end{figure}

The technical challenge in all investigations of nuclear
bound states is the accurate 
inclusion of all short-range correlations in the nuclear WF. 
Due to these short-range correlations, the partial wave decomposition
of nuclear WF's is very slowly converging. This hold
especially for the very tightly bound $\alpha$ particle 
WF. Therefore, a
rewriting of the Schr\"odinger equation for the 4N system 
\begin{equation}
H \ \Psi = \left( \ T + \sum_{i<j} V_{ij} + \sum_{i<j<k} V_{ijk} \
\right) 
\ \Psi = E \ \Psi
\end{equation}
according to the formalism of Yakubovsky \cite{yakubovsky67} 
is useful. We take NN pair potentials $V_{ij}$  and 3N potentials 
$V_{ijk}$ into account. $T$ denotes the kinetic energy operator, $H$ the full 
4N Hamiltonian and $\Psi$ the 4N WF. 
We will use Jacobi coordinates (see Fig.~\ref{fig:jacobi}) 
to represent our WF and dynamical 
equations. These separate the center of mass motion and, at the same time, guarantee 
a kinetic energy operator independent from angular variables. But these coordinates 
do not include { \it all} kinds of pair coordinates at the same time and it is hard 
to describe the short range correlations of pairs in other coordinates than their 
own relative coordinate. Other coordinates unavoidably lead to strong angular dependences 
or, in other words, to a very slowly converging series of partial waves. On the other hand, 
the Jacobi coordinates include the relative coordinates of {\it some} pairs. Correlations 
of those pairs are easily described. The WF contains the correlations 
of all pairs and is hard to expand in Jacobi coordinates. This makes the decomposition
of the WF in Yakubovsky components (YC) highly advisable. The YC's single 
out clusters of the four particles. The way they are defined guarantees that they are driven 
by correlations within these clusters only. Therefore, they are efficiently expanded 
in Jacobi coordinates, which single out the same clusters.     

In the isospin formalism nucleons are identical particles. 
This implies several symmetry properties, which connect
the different YC's and reduce the number of independent 
coupled equations and YC's to two. 
The following set of Yakubovsky  equations (YE's) are obtained   
for the two YC's $\psi_{1A}$ and $\psi_{2A}$ \cite{kamada92a,glockle93a,nogga01b}
\begin{eqnarray}
\label{eq:alphaeq1}
\psi_{1A} & \equiv & \psi_{(12)3,4} = G_0 t_{12} P  
\ [ (1-P_{34}) \ \psi_{1A} + \psi_{2A} ] + (1+G_0 t_{12}) \ G_0 \ V_{123}^{(3)} \ \Psi \\
\label{eq:alphaeq4}
\psi_{2A} & \equiv & \psi_{(12)34} = G_0 t_{12} 
                     \tilde P [ ( 1- P_{34}) \psi_{1A} + \psi_{2A} ] 
\end{eqnarray}
The other YC's are replaced by transposition operators $P_{ij}$ and combinations 
$P=P_{13}P_{23} + P_{12}P_{23}$
and $\tilde P = P_{13} P_{24}$, acting on the two 
remaining YC's. 
The kinetic energy enters  through the free propagator $G_0 = { 1 \over E - T }$
and the pair interaction by means of the pair $t$-matrix $t_{12}$. The 3NF's show up 
in the interaction term $V_{123}^{(3)}$. This defines a part of the 3NF in the cluster 
(123), which is symmetric in the pair (12) and which can be related by 
an interchange of the three particles to two other 
parts  $V_{123}^{(1)}$ and $V_{123}^{(2)}$ 
that sum up to the total 3NF of particles 1,2 and 3:
$ V_{123} = V_{123}^{(1)} + V_{123}^{(2)}  + V_{123}^{(3)}$.
For 3NF's based on a meson-exchange picture, 
$ V_{123}^{(3)}$ describes the interaction induced by a meson interchanged 
between particles 1 and 2 and, on the way, re-scattered by the third particle, as 
indicated in Fig.~\ref{fig:3nf}. 

Applying a combination of transpositions to the set of YC's, one obtains 
the WF as  
\begin{equation}
  \label{eq:alphawavef}
  \Psi = [1-(1+P)P_{34}] (1+P) \psi_{1A} + (1+P)(1+\tilde P) \psi_{2A}
\end{equation}
The YC's $\psi_{1A}$ and $\psi_{2A}$ 
are anti-symmetric in the pairs (12) or (12) and (34), 
respectively \cite{nogga01b}. 
This guarantees the total anti-symmetry of the WF $\Psi$.

The YC's are expanded in their ``natural'' Jacobi coordinates. This means that 
$\psi_{1A}$ is represented in the coordinates shown in the top 
of Fig.~\ref{fig:jacobi}, because both single out the pair (12) and the cluster (123). 
$\psi_{2A}$ singles out both pairs, (12) and (34), and is simplest, when expanded in 
the coordinates, shown at the bottom of Fig.~\ref{fig:jacobi}. 
Each of the coordinates involves three relative momenta 
$p_{12}$, $p_{3}$ and $q_{4}$ or  $p_{12}$, $p_{34}$ and $q$, respectively. 
The angular dependence is expanded in partial waves, leading to three  
orbital angular momentum quantum numbers for each kind of coordinate:
$l_{12}$, $l_{3}$ and $l_{4}$ or $l_{12}$, $l_{34}$ and $\lambda$. 
We use $jj$ coupling. Therefore, we couple, as indicated in the figure, 
the orbital angular momenta and corresponding spin quantum numbers  
to the intermediate quantum numbers $j_{12}$, $I_3$ and $I_4$ or $j_{12}$ and $j_{34}$,
and these are coupled to the total angular momentum $J$ and its third component $M$,
using two additional intermediate angular momenta $j_3$ and $I$:
$ (( j_{12} I_3 ) j_3  I_4 ) J M $ or $ (( j_{12} \lambda ) I j_{34}) JM $.
For the isospin quantum numbers (see Fig.~\ref{fig:jacobi}) similar 
coupling schemes to total isospin $TM_T$ involve 
only one intermediate quantum number $\tau$: 
$((t_{12} \ {1 \over 2 } ) \tau \ {1 \over 2 } ) T M_T $
or $(t_{12} t_{34}) T M_T $.

As we already pointed out, the partial wave decomposition requires a huge number of partial 
waves, whenever one needs to represent correlations of pairs and clusters in coordinates,
which don't single them out. Unfortunately, this is still necessary in the intermediate 
states in Eqs.~(\ref{eq:alphaeq1}) and (\ref{eq:alphaeq4}) (for $P \ \psi_{1A}$, etc.).
Therefore, we still need a tremendous number of partial waves to find converged results.
However, we are going to show numerically  in Section~\ref{sec:resb} that we can speed 
up the convergence greatly by using the Yakubovsky decomposition. 
For our calculations, we decided to truncate the orbital 
angular momenta,  requiring that $j_{ij} \le 6$ and $l_i, \lambda \le 8$.
Additionally, we constrain the expansion for both kinds of coordinates by another parameter 
$l_{sum}^{max}$, requiring that 
$l_{12}+l_{3}+l_{4} \le l_{sum}^{max}$  and $l_{12}+l_{34}+ \lambda \le l_{sum}^{max}$. 

We use $l_{sum}^{max}=14$.
Our most sophisticated calculations, including the 
$T=1$ and $T=2$ isospin channels, need  a total number of 
4200 partial waves for the first kind of coordinates and 2000 for the second kind. 
We require 36-40 mesh points to discretize the magnitudes of each of the momenta 
$p_{12}$, $p_{3}$ and $q_{4}$ or $p_{12}$,  $p_{34}$ and  $q$.
This insures that we obtain results for the binding energy of the 
$\alpha$ particle, which are converged within 50~keV.  
 
Using this partial wave truncation, we find that the discretized integral kernel 
for the set of Eqs.~(\ref{eq:alphaeq1}) 
and (\ref{eq:alphaeq4})  is of the dimension $3 \cdot 10^8 \times 3 \cdot 10^8$.  
Clearly this can no longer be treated by standard techniques of
numerical linear algebra, like the QR-algorithm, and one is forced
to use an iterative scheme. A Lanczos type method 
\cite{saake,stadler91}  has turned
out to be very powerful in the past and also here.  Succinctly, 
for an arbitrary $N$-component starting  vector  for the unknown
amplitude, one applies the kernel leading to
a new  vector. This is repeated  several times by applying the kernel
 always to the new vectors.  
That set of vectors is then orthonormalized and the unknown
amplitude expanded into those elements. 
Inserting this expansion  again into the
eigenvalue equation Eqs.~(\ref{eq:alphaeq1}) 
and (\ref{eq:alphaeq4}),  
one ends up with a small set of  linear algebraic
eigenvalue equations of dimension $n$, where $n$ counts the number of
applications of the kernel. $n$  is typically 10-20. 
The energy eigenvalue $E$, which is
buried as a parameter in the kernel is determined in such a manner that
the eigenvalue of the kernel is 1.

Another challenge is the application of the three-nucleon force. In
momentum space and partial-wave decomposed, this is a huge matrix of
typical dimension $5 \times 10^4 \times 5 \times 10^4$ 
for each total 3N angular momentum and parity.   
In case of the
$\alpha$ particle, the 3N subsystem total angular momenta have to be taken
into account up to $17 \over 2$. 
Instead of  preparing  these matrices, we handle the
3N forces differently. They can be naturally broken up into a sequence of
pseudo two-body forces with a change of Jacobi momenta in between 
(transpositions). This has been described, for  the first time in \cite{huber97b}. 
The generalization to the 4N system is described 
in Appendix~\ref{app:3nf}.
This technique is much more efficient and  
even allows one to evaluate the 3NF's in
each new iteration of the kernel -- no storage of huge intermediate
matrices related to 3NF's is required.

A typical run on a massively  parallel T3E 
with  128  processors takes 2~hrs to get one eigenvalue and the corresponding 
eigenvector.  For our
method of parallelization  we refer to \cite{noggaphd}.

\section{Results}

\subsection{Adjustment of 3NF's}
\label{sec:resa}

In this paper we restrict ourselves to the modern realistic 
NN interactions, which are all fitted to the NN data with the same 
high accuracy and also provide a nn force, which predicts 
a reasonable nn scattering length. These interactions are the AV18 
\cite{av18} and the CD-Bonn \cite{cdbonn}. Additionally we 
show results for the Nijm~I, Nijm~II and Nijm~93 interactions \cite{nijm93},
which are not adjusted to the nn scattering length and in 
case of Nijm~93 give a slightly less accurate fit to the NN data. 
The results for the $^3$He and $^3$H BE's are shown in 
Table~\ref{tab:3nbound}. They are based on calculations, which 
take two-body angular momenta up to $j_{12}=6$ into account and are converged 
up to 2~keV. The full charge dependence of the interaction as well as 
the n-p mass difference are considered. Also the Coulomb force is 
included exactly as described in \cite{noggaphd,nogga00}.    

\begin{table}

  \begin{center}
    \begin{tabular}[t]{l|rr|rr|r}
  & \multicolumn{2}{c|}{$^3$H} & \multicolumn{2}{c|}{$^3$He} &    \cr
interaction       &  $E_B$  &    $T$   &  $E_B$  &    $T$  & $\Delta E_B$ \cr
\hline
CD-Bonn           & -8.013  &  37.43   & -7.288  & 36.62   &  0.725       \cr
AV18              & -7.628  &  46.76   & -6.917  & 45.69   &  0.711       \cr
\hline
Nijm~I            & -7.741  &  40.74   & -7.083  & 40.01   &  0.658       \cr
Nijm~II           & -7.659  &  47.55   & -7.008  & 46.67   &  0.651       \cr
Nijm~93           & -7.668  &  45.65   & -7.014  & 44.79   &  0.654       \cr
\hline
Exp.              & -8.482  &  ---     & -7.718  & ---     &  0.764   \cr
    \end{tabular}
    \caption{3N binding energies $E_B$ for different NN interactions 
             compared to the experimental values. 
             Results are shown for $^3$H,$^3$He and their binding 
             energy difference $\Delta E_B$. 
             Additionally, we show the kinetic energies $T$. 
             All results are given in MeV}
    \label{tab:3nbound}
  \end{center}
\end{table}

As well known \cite{stadler91,sauer86,friar93,wu93,kievsky95} 
all NN model interactions lead to an underpredicted 3N BE. 
The underprediction is strongly model dependent and 
ranges from 0.8~MeV to 0.5~MeV 
for the most modern interactions (see Table~\ref{tab:3nbound})
though their description of the NN data is comparable. For 
benchmark purposes, we also show results for the 
expectation value of the kinetic energy. These tend to be smaller 
for the non-local interaction Nijm~I and CD-Bonn. This behavior 
can be traced back to the softer repulsive core 
of non-local NN forces. We also show the binding energy 
difference of the two mirror nuclei $\Delta E_B$. One sees that 
all models underpredict the experimental value. The deviation
is somehow larger for the Nijmegen interactions, which do not 
describe the nn scattering length correctly. The additional 
differences for the Nijmegen interactions are, therefore, likely 
a result of an inadequate description of the NN scattering data.
We will address the issue of the $^3$H-$^3$He 
binding energy difference in Ref.~\cite{nogga01c}; 
therefore, we do not want to 
go in details here. 

Two possible dynamical ingredients are still missing in our calculations:
relativistic effects and 3NF's. We will not address the interesting 
question of including relativity in few-nucleon dynamics here. 
Attempts to understand this issue can be found 
in Refs.~\cite{rupp92,sammarruca92,stadler97a,stadler97b,glockle86,forest99}.
The results of those calculations are varying. 
Whereas approaches based on field equations,
like Bethe-Salpeter or Gross equations generally predict an increased 
binding energy compared to the non-relativistic solution, the calculations 
based on a relativistic Schr\"odinger equation predict a decreased 
binding energy. In the latter case the relativistic effects
are driven by boost properties, whereas in field theoretical 
approaches additional dynamical effects also occur.
The magnitude of the predicted effects is of the order 
of 200~keV.  
The problem is not yet solved. It has also been observed that 
relativistic effects and 3NF effects are related and cannot 
be separated in field equation approaches \cite{stadler97b}. 
In this paper we neglect all relativistic effects, hoping 
that part of them are included in effective 3NF terms. 

The knowledge regarding 3NF's is similarly scarce, as for the relativistic 
effects. It has been shown in Ref.~\cite{polyzou90} that
3NF's are not defined independently from the accompanying
NN interactions. Two 3N Hamiltonians based on two different, but
phase equivalent NN interactions, can be augmented by a properly
chosen 3N interaction to be equivalent in the 3N system.
In \cite{nogga00} we formulated the more inclusive statement
that one could, in principle, always find NN interactions, which replace a
3N interaction completely in a 3N Hamiltonian.
Ref. \cite{polyzou90} does not conclude that this is {\it always}  possible.
It is clear anyhow that the transformation are complicated and
therefore it is not practicable to use them to get rid of the 3NF's.
As soon as one includes relativistic features the 
Poincar{\'e} algebra inevitably enforces 3NF's 
\cite{foldy74}, which cannot be transformed away.

In view of this connection of 3NF models and NN force models,
a phenomenological approach to the 3NF is justified: 
given a 3NF model, one adjusts its parameters 
in conjunction with one NN interaction model to 3N or other 
nuclear data leading to different parameter sets of the 3NF 
for each NN interaction.

For the Urb-IX 3NF the parameters have been fixed in 
conjunction with the AV18 interaction using the $^3$H BE and the nuclear 
matter density predicted by this combination \cite{pudliner97}.
The TM force originally has not been adjusted in this way. 
Its parameters have been deduced from model assumptions 
and using $\pi$N scattering data \cite{coon79,coon93,coon01}.
It is clear that a complete 3NF based on meson-exchange 
should include not only $\pi$-$\pi$, but also 
  $\rho$-$\pi$, $\rho$-$\rho$ and so on exchanges. 
Attempts to include these processes have been done, but 
conclusive results, fixing the parameter sets,  
could not be obtained \cite{stadler95}. Therefore, we assume 
in our study that we can effectively include the effects of heavier 
mesons in the $\pi$-$\pi$ exchange TM model by a variation of the 
$\pi$NN form factor parameter $\Lambda$. It has been observed 
\cite{mckellar84,stadler91} that the $^3$H BE is sensitive to this cut-off. 
The original value $\Lambda=5.8 m_\pi$ has been fixed 
by matching the Goldberger Treiman discrepancy \cite{coon01}.
However, as has been argued in \cite{robilotta86}, 
the form factors are ill-defined, because they strongly influence 
the long-range part of the 3NF. Therefore an adjustment is justified. 
We emphasize that the aim of this paper is the investigation 
of model dependences due to the different 3NF's. 
To this aim we only require 3NF models, which are different and 
which have a sufficiently rich spin-isospin dependence.  
An adjustment of the 3NF does not spoil these requirements. 

We combined in Refs.~\cite{nogga97,nogga00} the available 
NN interactions with the TM 3NF and tuned $\Lambda$ 
to reproduce the $^3$H or  $^3$He BE's. 
The resulting $\Lambda$ values are shown in Table~\ref{tab:3nbind3nf}.
The table also includes results for a modified TM interaction. 
It has been argued in \cite{friar98} that the 
long-range/short-range part of the $c$-term is not consistent 
with chiral symmetry. Dropping it leads to a changed set of 
parameters, which we refer to as TM'. The parameters of TM and TM' 
are summarized in Table~\ref{tab:3nfpara} of Appendix~\ref{app:3nf}. 
    
\begin{table}

  \begin{center}
    \begin{tabular}{l|r|rr|r}
interaction  &  $\Lambda$ & $E$($^3$H)   & $E$($^3$He)   & $\Delta E_B$  \cr
\hline
CD-Bonn+TM   &   4.784    &  -8.478      & -7.735       & 0.743     \cr
AV18+TM      &   5.156    &  -8.478      & -7.733       & 0.744     \cr
AV18+TM'     &   4.756    &  -8.448      & -7.706       & 0.742     \cr
AV18+Urb-IX  &  ---       &  -8.484      & -7.739       & 0.745    \cr
AV18+Urb-IX 
 (Pisa) \cite{kievskypriv}
             &   ---      &  -8.485      & -7.742       & 0.743    \cr
AV18+Urb-IX 
 (Argonne) \cite{wiringa00}
             &   ---      &  -8.47(1)    & ---          &  ---      \cr
\hline
Exp.         &   ---      &  -8.482      & -7.718       &  0.764    \cr  
    \end{tabular}
    \caption{3N binding energy results for different combinations 
             of NN and 3N interactions, together with the adjusted form
             factor parameters $\Lambda$ in units of $m_\pi$. 
             The binding energies for $^3$H 
             $E$($^3$H)  and $^3$He $E$($^3$He) are shown. For completeness
             the splitting $\Delta E_B$ is also displayed. All energies are given 
             in MeV.}
    \label{tab:3nbind3nf}
  \end{center}
\end{table}

The fits have been done using less accurate 
BE calculations not including the isospin $T={3 \over 2}$ 
component and not including the effect of the n-p mass difference. 
Therefore, the new results for the BE's, shown in the table, do not exactly match 
the experimental values. The deviations are non-significant for 
the following study, so we refrain from refitting 
the $\Lambda$'s. We adjusted TM to the $^3$H BE and  
TM' to the $^3$He BE. 
The table also shows our results using the Urb-IX 
interaction, as defined in \cite{pudliner97}. 

Table~\ref{tab:3nbind3nf} confirms at the same time a well-known 
scaling behavior of the Coulomb interaction 
with the BE of $^3$He \cite{friar87}. 
The adjusted 3N Hamiltonians predict very similar 3N binding energies
and $\Delta E_B$'s.
This removes the model dependence of $\Delta E_B$ 
found in Table~\ref{tab:3nbound}.  We observe that the model 
independent prediction for these energy difference deviates from 
the experimental value by about 20~keV. Again, we refer to 
Ref.~\cite{nogga01c} for a more detailed discussion of this issue. 
In the same reference, a detailed comparison with 
hyper-spherical variational calculations is given. 
In Table~\ref{tab:3nbound}, for comparison, we only show  
the BE's obtained by the Pisa and Argonne group. 
We note that the calculation 
by the Pisa group is in full agreement with our results. 
The small deviation from the Argonne result is not significant 
in view of the comparably large statistical error bar 
of the GFMC calculation.  

We are now ready to apply the 3N model Hamiltonians, 
given by the $\Lambda$ values in Table~\ref{tab:3nbind3nf},
to the 4N system. By using the models from Table~\ref{tab:3nbind3nf}, 
we insure that dependences due to scaling effects, as visible 
for example in $\Delta E_B$, are excluded. Given the very different 
functional forms of the Urb-IX, TM and TM' interactions, we 
can expect to see any remaining model dependences in our 
calculations.

\subsection{$\alpha$ particle binding energies}
\label{sec:resb}

Based on these model Hamiltonians, we solved the YE's~(\ref{eq:alphaeq1}) 
and (\ref{eq:alphaeq4}) with no uncontrolled approximation. 
The following results are based on a partial 
wave decomposition truncated using $l_{sum}^{max}=14$. It has been 
verified that this is sufficient to obtain converged BE's 
with an accuracy of 50~keV. The binding energies given were found 
varying the energy parameter in Eqs.~(\ref{eq:alphaeq1}) 
and (\ref{eq:alphaeq4}) until the eigenvalue 1 appears 
in the spectrum of the set of YE's. 

Independently, one can check the results with a calculation of 
the expectation value of the Hamiltonian. We emphasize 
that this is  an important feature of our method, which minimizes 
the possibility of errors in the codes or unexpected numerical 
difficulties. 

For these checks one faces the problem to represent 
the WF with high accuracy. We  already pointed out
that the WF of the $\alpha$ particle is 
extremely slowly converging, 
because there is no set of Jacobi momenta suitable 
to describe the short range correlation in {\it all} NN pairs.
In Table~\ref{tab:av18conv} we exemplify the convergence 
behavior of the WF for the AV18 interaction. 
The normalization and the expectation values of the kinetic energy, potential 
energy and Hamiltonian are shown. The WF's have been derived
from the same set of YC's, using Eq.~(\ref{eq:alphawavef}). 
The calculation of the WF is based on a partial wave decomposition 
truncated with $l_{sum}^{max}=14$. In this way we obtained the WF 
in the two different representations, depicted in Fig.~\ref{fig:jacobi}.  
For the expectation values shown in the table, we truncated 
the WF in a second step to the partial waves 
given by the $l_{sum}^{max}$ parameter in the first column. 
It turned out that the evaluation of the kinetic 
energy is difficult, because $T$ amplifies the slowly converging 
high momentum components of the WF. The kinetic-energy
expectation values, shown in the fourth and fifth columns of the table, 
do not converge within the chosen partial-wave truncation. 
However, one can rewrite the kinetic energy using Eq.~(\ref{eq:alphawavef}) 
and the fact that the transposition operators  commute with the kinetic 
energy and simply result in a sign change, 
if applied to a fully anti-symmetrized WF:
\begin{equation}
\label{eq:kintrick}
<\Psi | T | \Psi > = 12 <\Psi | T | \psi_{1A} > + 6 <\Psi | T | \psi_{2A} >
\end{equation}
The right-hand side involves mixed matrix elements with the YC's. 
The first term has to be evaluated in the 1A representation, because $\psi_{1A}$
is given in these coordinates, and the second term in the  2A ones 
because of the coordinates 
of $\psi_{2A}$. The results for $T$ based on this equation are shown 
in the column labeled $T(mix)$ and show a promising convergence 
behavior. We observe a much faster convergence for the  YC's, which 
was expected and which justifies the YE's approach to the 4N 
Schr\"odinger equation. Based on this experience, we normalize 
our WF and the YC's using a similar formula for the norm. 
Consequently, the deviation of directly calculated norms of the WF, shown 
in columns  2 and 3, from one is a measure of the numerical error 
of our anti-symmetization of the full WF. 

\begin{table}
\begin{center}
\begin{tabular}[t]{c|cc|ccc|cc|cc}
$l_{sum}^{max}$ & $<\Psi | \Psi > ^{1A}$ & $<\Psi | \Psi > ^{2A}$ 
& $<\Psi | T | \Psi > ^{1A}$ & $<\Psi | T | \Psi > ^{2A}$
& $T(mix)$ & $<\Psi | V | \Psi > ^{1A}$ & $<\Psi | V | \Psi > ^{2A}$
& $<H>^{1A}$ & $<H>^{2A}$ \cr
\hline
2   & 0.9117 & 0.9084 & 61.27 & 62.14 & 91.80 & -110.20 & -110.44 &-18.40 &-18.65 \cr
4   & 0.9662 & 0.9582 & 79.10 & 76.11 & 96.85 & -117.55 & -118.12 &-20.70 &-21.27 \cr
6   & 0.9820 & 0.9766 & 86.36 & 83.49 & 97.56 & -120.66 & -120.71 &-23.09 &-23.15 \cr
8   & 0.9927 & 0.9890 & 92.41 & 90.09 & 97.75 & -121.43 & -121.39 &-23.67 &-23.63 \cr
10  & 0.9961 & 0.9939 & 94.59 & 92.93 & 97.79 & -121.84 & -121.84 &-24.05 &-24.05 \cr
12  & 0.9982 & 0.9969 & 96.10 & 95.04 & 97.80 & -121.97 & -121.96 &-24.16 &-24.16 \cr
14  & 0.9990 & 0.9986 & 96.51 & 95.70 & 97.80 & -122.03 & -122.01 &-24.23 &-24.21 \cr
\end{tabular}
\end{center}

\caption{Convergence of the $\alpha$ particle WF for different truncations 
of the basis states. The superscripts $1A$ and $2A$ indicate 
the type of Jacobi coordinates employed. The results are based on a calculation
using the AV18 NN interaction and no 3NF. See text for details.}
\label{tab:av18conv}

\end{table}

Unfortunately, a similar approach is not possible for the expectation values 
of the potential. However, the interaction does not overemphasize 
the high-momentum tail 
and its expectation value is  much faster converging. 
We find a reasonable agreement of 0.02~\% of the 1A and 2A results
and convergence of both values to uncertainty of 60~keV. For completeness we show 
the expectation value of the Hamiltonian 
based on $T(mix)$ and the 1A or 2A expectation value of $V$. These 
values agree within 0.1~\%. The expectation values differ from the 
binding energy result of $-24.25$~MeV by only 20 to 40~keV.
This is well within the error of 60~keV, which has to be expected from the 
convergence behavior of $V$ and verifies the accuracy of our results. 
In the following, we will only present the binding energies,
the $T(mix)$ values and the expectation values of $H$ and $V$  
based on the 1A representation. We consider it more accurate than the $2A$ 
representation,
because the norm is closer to one.   

In Table~\ref{tab:alphabindnn} our $\alpha$ particle binding energies 
are summarized for Hamiltonians based on NN forces only. 
The results are identical to the ones published in 
Ref.~\cite{nogga00} except for AV18, where we present 
a new calculation, based on a more accurate grid and taking $T=1$ and $T=2$ components 
into account. Due to the more accurate momentum grid, our binding energy 
changed by 30~keV, well within our estimated numerical error
of 50~keV. Therefore, we did not redo the calculation 
for the other interactions. The table 
also shows a result obtained using the NCSM approach \cite{navratil00}.
Our result agrees with their 
number within the numerical errors estimated.  

\begin{table}
  \begin{center}
    \begin{tabular}{l|rrrr}
 interaction &  $E_\alpha$  & $H$       & $T$       &  $V_{NN}$       \cr
\hline                                                          
Nijm~93        & -24.53     & -24.55    &  95.34    & -119.89   \cr
Nijm~I         & -24.98     & -24.99    &  84.19    & -109.19   \cr
Nijm~II        & -24.56     & -24.55    & 100.31    & -124.86   \cr
AV18           & -24.25     & -24.23    &  97.80    & -122.03   \cr
CD-Bonn        & -26.26     & -26.23    &  77.15    & -103.38   \cr
\hline                                                          
CD-Bonn                                                         
 \cite{navratil00}                                              
               & -26.4(2)   &   ---     & ---       & ---       \cr
\hline                                                          
Exp.           & -28.30     &   ---     &  ---      & ---       \cr
    \end{tabular}

    \caption{$\alpha$ particle binding energy predictions $E_\alpha$
             of several NN potential models compared to the experimental value 
             and the ``no-core shell model'' result \cite{navratil00}. 
             The expectation values of the kinetic energy $T$, the 
             NN interaction $V_{NN}$ and 
             the Hamiltonian operator $H$ are also shown.  
             All energies are given in MeV. }
    \label{tab:alphabindnn}
  \end{center}
\end{table}

As in the case of the 3N BE's, the 4N BE's are also underpredicted 
by all modern NN force models.The underbinding ranges from 2 to 4~MeV,
showing that the results are also strongly model 
dependent. Once again the non-local forces predict more  
binding and, similarily, a reduced kinetic energy. 
The expectation values of $H$ agree within the numerical accuracy 
of 50~keV with the BE's $E_\alpha$, which have been directly 
obtained from the YE's. 

\begin{figure}
\begin{center}
\psfrag{xxx}{\Large $E$($^3$H) [MeV] }
\psfrag{yyy}{\Large $E$($^4$He) [MeV] }
\includegraphics[angle=0,width=7cm]{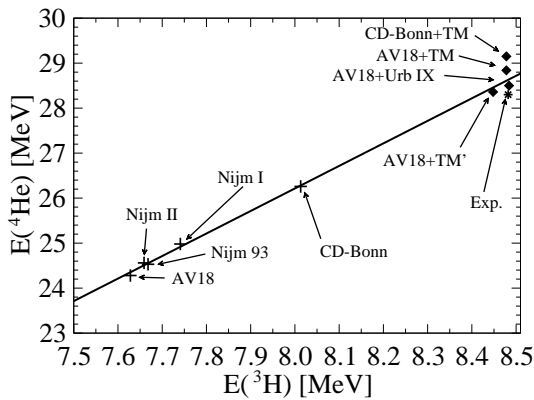}	
\end{center}

\caption{Tjon-line: $\alpha$ particle binding-energy predictions $E$($^4$He) 
         dependent on the predictions for the $^3$H binding energies 
         for several realistic interaction models.
         Predictions of interaction models  
         without (crosses) and with (diamonds)  a 3NF
         are shown. The experimental point is marked by a star. The line
         represents a least square fit to the predictions of models 
         without a 3NF. }

\label{fig:tjonline}

\end{figure}

In Ref.~\cite{tjon75} an fascinating linear correlation 
of the $\alpha$ particle and $^3$H BE's has been observed,
known as the Tjon-line. 
Our new results confirm this correlation for the newest
NN forces. This is displayed in Fig.~\ref{fig:tjonline}. 
One sees that all predictions based on only NN forces are situated 
on a straight line. However, the experimental point slightly deviates 
from this line hinting to dynamical ingredients 
beyond the NN interaction and the non-relativistic Schr\"odinger
equation. We also observe a strong dependence of this result 
on the accuracy of the NN force. Omitting the electromagnetic 
part of the AV18 NN interaction leads to 16~keV overbinding 
for the deuteron. A calculation based on this potential resulted 
in a visible deviation from the Tjon-line.

In the next step we also include 3NF's into our Hamiltonian. 
As discussed above, we adjusted these force in conjunction 
with the different NN interactions. We expect 
a much smaller dependence of the BE's on the 
3N Hamiltonians in this case, because 
we remove in this way model dependences, which are correlated 
to the 3N BE. As one learns from the Tjon-line
these are the dominant ones. 
Our results are given in Table~\ref{tab:alphabind3nf}. 
Again we obtained an accuracy of the BE's $E_\alpha$ of 50~keV. 
The convergence is slower for these calculations. Therefore, 
we do not find the same accuracy for the expectation values
as for the BE's. 
For these we estimate an error of 100~keV, which is still 
within 0.2~\% of the kinetic energy. 

\begin{table}

  \begin{center}
    \begin{tabular}{l|rrrrr}
 interaction &  $E_\alpha$  & $H$       & $T$       &  $V_{NN}$ & $V_{3NF}$ \cr
\hline                                                          
CD-Bonn+TM     & -29.15     & -29.09    &  83.92    & -106.16   & -6.854   \cr
AV18+TM        & -28.84     & -28.81    & 111.84    & -132.62   & -8.033   \cr
AV18+TM'       & -28.36     & -28.40    & 110.14    & -133.36   & -5.178   \cr
AV18+Urb-IX    & -28.50     & -28.53    & 113.21    & -135.81   & -5.929   \cr
\hline
AV18+Urb-IX 
(Argonne) \cite{wiringa00}
               & -28.34(4)  &   ---     & 110.7(7)  & -135.3(7) & -6.3(1)  \cr
\hline
Exp.           & -28.30     &   ---     &  ---      & ---       &   ---    \cr
    \end{tabular}

    \caption{$\alpha$ particle binding-energy predictions $E_\alpha$
             for the CD-Bonn and AV18 interactions in conjunction with various  
             3NF's, compared to the experimental value and the 
             Argonne-Los~Alamos result. 
             The expectation values of the kinetic energy $T$, the 
             NN interaction $V_{NN}$, the 3NF $V_{3NF}$ and 
             the Hamiltonian operator $H$ are also shown.  
             All energies are given in MeV. }
    \label{tab:alphabind3nf}
  \end{center}
\end{table}

For the NN and 3N forces used, we observe a small 
overbinding of 60 to 800~keV. These results are also included 
in Fig.~\ref{fig:tjonline}. For the TM' and Urb-IX results 
we find only small deviations of our results from the 
Tjon-line. For TM we see more deviations. The TM force 
seems to destroy the correlation between the 
$^3$H and $\alpha$ particle 
BE's. Though the TM' force and the Urb-IX interaction 
are quite different, their BE predictions seem to be comparable. 
Unfortunately the adjustment of the 3N force has not been 
done with the same accuracy for TM'. In view of the very 
expensive calculations necessary to improve the 
TM' results and in view of the expected agreement of the 
TM' and Urb-IX BE's, we did not re-calculate 
for TM', but omit its results in the following argumentation. 
The average BE for the $\alpha$ particle using only a NN 
interaction (based on the restricted choice shown in 
Table~\ref{tab:alphabindnn}) is -24.9~MeV or 88~\% of the 
$\alpha$ particle BE. Based on the TM and Urb-IX 
results in Table~\ref{tab:alphabind3nf}, we estimate 
an average 3NF contribution to the $\alpha$ particle binding 
of 3.9~MeV or  14~\% of the experimental BE. From the 
same results we find an average overbinding of 500~keV or 
2~\% of the BE. The contribution of the 3NF is strongly 
dependent on the NN interaction due to the adjustment 
of these forces to the 3N BE. The model dependence 
of the overbinding is much smaller, but depends 
on the NN and the 3N force.  
One can consider this overbinding 
as the effect of a missing repulsive 4N force. The average size 
of this force can be expected to be 2~\% of the BE
in the $\alpha$ particle. Certainly the size of this force 
will be related to the NN and 3N forces used. 
The approach employed in \cite{polyzou90} shows that these 4N
forces  are related to the 3N Hamiltonians in the same way as the 
3N forces to the 2N Hamiltonians. We conclude from our results 
that we have found numerical evidence that 4N forces are, indeed, 
much smaller than 3N forces, at least in conjunction with 
today's NN and 3N interactions.  We do not exclude that 
new additional 3NF terms could be found,
which reduce the necessary contribution of 4N forces. 
The results  support the generally accepted 
assumption that meaningful nuclear-structure calculations can be performed 
utilizing bare NN and 3N interactions in a  
microscopically  self-consistent manner.
 We expect that 4N forces 
probably show up in heavier nuclei in the same order 
of magnitude (2~\% of the BE). We, therefore, suggest to take 
an error of this size into account, when one discusses 
BE's for systems with $A>4$, based on present NN and 3N forces.

\subsection{Properties of the $\alpha$ particle WF}
\label{sec:resc}

Besides the BE's, we are also interested in the WF 
of the 4N system, because it serves as  input to several 
analyses of experiments involving the $\alpha$ particle. 
Most of these calculations are based on plane wave impulse
approximation (PWIA). These calculations are directly sensitive 
to the WF. Model dependences of the WF are hints to model 
dependences of these observables. However, because WF's are not 
observable themselves, we emphasize that these dependences might 
disappear once the full dynamics are taken into account.

We start with a contribution of the different isospin states 
to the WF. Because we made a full, isospin breaking 
calculation for AV18 only, there is only one result shown 
in Table~\ref{tab:isoprops}. The results for the 3N system
do not depend on the interaction used \cite{nogga01c}. 
Therefore, we do not expect model dependences here. 
 
\begin{table}

\begin{center}
\begin{tabular}{l|rrr}
interaction &  $T=0$ &  $T=1$ &  $T=2$ \cr
\hline
AV18       & 99.992 &  0.003 &  0.005  
\end{tabular}
\end{center}
\caption{Contribution of different total isospin states 
         to the $\alpha$ particle wave function. 
         The values are given in \%.}
\label{tab:isoprops}

\end{table}

One sees an extremely small contribution of the $T=1$ 
and $T=2$ component to the WF. However, 
it is of interest that our $T=1$ probability, based on realistic 
nuclear forces, is larger than the one estimated in \cite{ramavataram94}. 
There the $T=1$ admixture has been found to be about $7\cdot 10 ^ {-4}$~\%
and the $T=2$ state has not been considered. 
We found the 
$T=2$ component nearly twice as large as 
the $T=1$ admixture.
Moreover the form of our $T=1$ state will also be different from the one 
in \cite{ramavataram94}. 
As a consequence the isospin admixture correction to the asymmetry as given in 
\cite{ramavataram94} will change. 
A renewed evaluation of that correction, also 
including the larger $T=2$ state has not been carried through, 
but appears interesting  in view of ongoing experiments.

WF properties are also important for comparisons to other 
calculational schemes for treating the 4N system. 
Among the most simple of these properties
are the $S$, $P$ and $D$-wave probabilities 
of the WF. These are given in Table~\ref{tab:spdalpha}
for the models based on the CD-Bonn and AV18 interactions. 
The values given in Table~\ref{tab:spdalpha} are based on overlaps 
between the YC's and the WF's, similar to those for the kinetic 
energies. These numbers are more accurate than the results 
given in \cite{nogga00}. However, the differences are 
not significant, as they affect only the last digit 
of the results. 

As expected the orbital $S$-state is dominant. The $D$-state 
probability is sizable, very similar to results for $^3$H 
\cite{nogga01c}, and the $P$-state gives only a small  
contribution. The $D$-state probability for 3N Hamiltonians, based 
on CD-Bonn, is smaller than those for models based on the AV18. 
This is related to the smaller tensor force of non-local 
interactions. It sticks out that all 3NF's lead to an 
increase of the $P$-wave probability by a factor of 2.

\begin{table}[tp]
  \begin{center}
    \begin{tabular}{l|rrr}
interaction     &  $S$       &    $P$    &     $D$     \cr
\hline                                                
CD-Bonn         & 89.06      &  0.22     &  10.72       \cr
CD-Bonn+TM      & 89.65      &  0.45     &   9.90      \cr
AV18            & 85.87      &  0.35     &  13.78      \cr
AV18+TM         & 85.36      &  0.77     &  13.88      \cr
AV18+TM'        & 83.58      &  0.75     &  15.67      \cr
AV18+Urb-IX     & 83.23      &  0.75     &  16.03      \cr
    \end{tabular}
    \caption{$S$, $P$ and $D$ state probabilities for the $^4$He
      wave functions. All probabilities are in \%.   }
    \label{tab:spdalpha}
  \end{center}
\end{table}

This raises the question, whether the 
considered 3NF really act differently in the 4N system. Because 
of the scarce knowledge on 3NF's, this issue is very important.
It insures that we get insight into possible 
impacts of 3NF's in general, only if our models cover 
a wide range of interactions. To verify this issue, 
we decompose the WF's into parts with different 
total orbital angular momentum, namely $S$, $P$ and $D$-states. 
Based on these components, we calculate the expectation 
values of the Urb-IX and TM 3NF's for three different 
WF's. One is based on the AV18 interaction only, one on 
the AV18+Urb-IX and the last on the AV18+TM. 
Four kinds of matrix elements dominate the total 
expectation value of the 3NF: the diagonal $S$-$S$-state
and $D$-$D$-state matrix elements and the overlaps 
of $S$-state with $P$- and $D$-state. 
Table~\ref{tab:orbang3nfexp} shows our results. 
In the second and fourth columns expectation values for 
Urb-IX and TM are shown for the same WF, based on AV18. 
One observe a strong disagreement of these matrix elements. 
The diagonal elements for Urb-IX are strongly 
repulsive. They seem to be driven by the isospin and spin 
independent, phenomenological short-range core of the 
Urb-IX model. In strong contrast, the $S$-$S$ matrix 
element contributes most of the attraction in the case of the TM. 
The attraction of Urb-IX is contributed by the $S$-$D$ 
overlap. This is a major difference 
in the action  of both models in the 4N system. 
It insures that we used, indeed, very different 3NF models
though both are based on the 2$\pi$ exchange mechanism. 
Additionally, we see in columns three and five of the 
table the expectation values based on WF's for the 
full Hamiltonian. These expectation values differ 
sizably from the ones based on the AV18 WF. We confirm 
for both 3NF's that a perturbative treatment of them  
is impossible. For the 3N system this was already 
emphasized in 
Refs.~\cite{friar88a,sasakawa86,bomelburg86}.
Especially interesting is the $S$-$D$ overlap 
of the TM force. The AV18 WF result is strongly 
repulsive, whereas the full calculation leads 
to a slightly attractive contribution. This suggests
interesting changes of the 3N configurations 
in the $\alpha$ particle due to this force. 

\begin{table}[tp]
  \begin{center}
    \begin{tabular}{l|r|r|r|r}
3NF        & \multicolumn{2}{c|}{Urb-IX} & \multicolumn{2}{c}{TM} \cr
\hline
WF         &  AV18    &   AV18+Urb-IX    &    AV18    &   AV18+TM   \cr 
\hline
 S-S       &   3.16   &   2.74           &   -2.34    &  -4.09      \cr
 S-P/P-S   &  -0.96   &  -2.10           &   -1.22    &  -3.56      \cr
 S-D/D-S   &  -5.44   &  -7.46           &    2.08    &  -0.14      \cr
 D-D       &   0.59   &   0.85           &    0.01    &   0.06      \cr
    \end{tabular}
    \caption{Contribution of different total 
             orbital angular momenta in the wave functions to the expectation 
             values of the Urb-IX and TM 3NF's. All energies are given in MeV.}
    \label{tab:orbang3nfexp}
  \end{center}
\end{table}

\begin{figure}
\begin{center}
\includegraphics[angle=0,width=7cm]{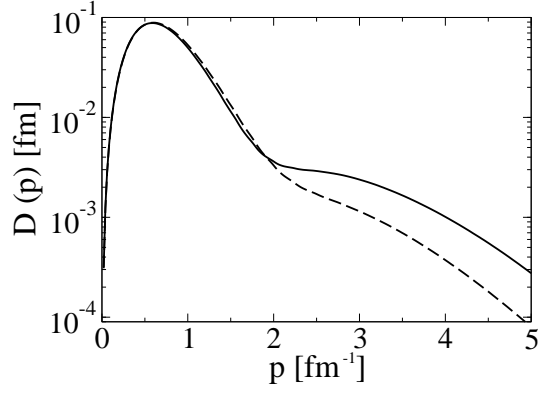}	
\end{center}

\caption{Nucleon momentum distributions        
         in $^4$He on a logarithmic scale.   
         The distribution functions  are based on 
         calculations using the AV18+TM (solid line) and 
         CD-Bonn+TM (dashed line)  
         potentials.
         The functions are normalized to $\int D(p) dp = { 1 \over 4 \pi } $.}
\label{fig:momnn}
\end{figure}

Are these changes in the 3N configuration visible in momentum 
distributions? We start in Fig.~\ref{fig:momnn} with a comparison 
of the nucleon momentum distribution 
\begin{equation}
  \label{eq:distrfunc}
  D(p) = {1 \over 4 \pi} \ 
    < \Psi \ J=0 \ M=0 | \  \delta(p-q_{4}) \ | \Psi \ J=0 \ M=0 >  
\end{equation}
for WF's based on different NN interactions. The momentum distributions
are angular independent. We only consider the $T=0$ 
components here. Therefore, the proton and neutron distributions 
are equal. Because we include in both calculations a 
3NF, the WF's are the result of calculations, which 
roughly give the same BE's. This insures that we do not 
find differences, which can be traced back to a higher density 
of the nucleus.
The distributions are equal for momenta below 
$p=1$~fm$^{-1}$ for both WF's. For momenta between $p=1$~fm$^{-1}$ 
and $p=2$~fm$^{-1}$ the deviations are moderate. 
Above this momentum the AV18 WF  is much bigger. 
We find a clear difference between CD-Bonn and AV18 in this momentum 
region.

\begin{figure}
\begin{center}

\includegraphics[angle=0,width=7cm]{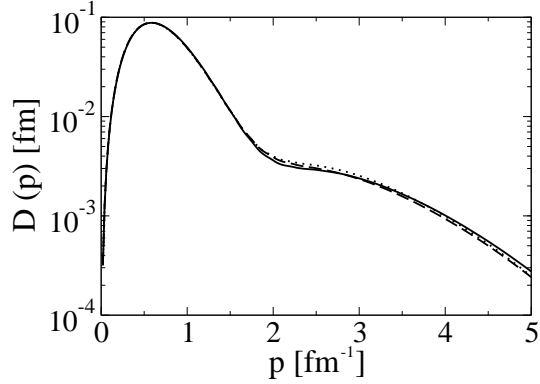}	

\end{center}

\caption{Nucleon momentum distributions        
         in $^4$He on a logarithmic scale.   
         The distribution functions  are based on 
         calculations using the AV18+TM (solid line), 
         AV18+Urb-IX (dotted line) and AV18+TM' (dashed line) 
         potentials.
         The functions are normalized to $\int D(p) dp = { 1 \over 4 \pi } $.}
\label{fig:mom3nf}

\end{figure}

We do not see similar deviations comparing the 
momentum distributions for different 3NF's. This is 
shown in Fig.~\ref{fig:mom3nf}. The WF's shown there are based on
the same NN interaction, AV18, but differ in the 3NF used. Again 
the BE's are comparable and there can be no deviations expected 
because of density differences. In fact one observes that 
the distributions are nearly equal for all models 
in the whole momentum range. This indicates a remarkable 
stability of momentum distributions with respect to the 
3NF. This is in accord with the same independence of 
the 3NF choice for $T$, the second moment of the
momentum distribution, 
as shown in Table~\ref{tab:alphabind3nf}.   

The correlations of two nucleons in nuclei are of 
great theoretical interest. Defined as the probability
that two nucleons have a certain distance 
inside the nucleus, one finds very similar 
correlations for nuclei with different $A$
\cite{forest96,noggaphd}. The correlation is characterized by the strong short-range 
repulsion of nuclear forces, leading to a small probability 
that two nucleons are close to each other. However, 
the quantitative results depend on the force model used. 
For the 3N system this has been shown in \cite{nogga97},
and we find similar results for the 4N system \cite{noggaphd}. 
These correlations are not observable. Therefore, 
difference in this WF property might not show up in 
observables. Electron induced scattering experiments, which 
intent to see these correlations also see effects 
of meson exchange currents (MEC's) and final state interactions (FSI's). 
Therefore a complete dynamical description of these 
processes is necessary \cite{glockle01a}.

Nevertheless, we want to show those correlations here. 
Two-nucleon  knock-out experiments are expected 
to provide information on relative momentum distributions 
(see for instance \cite{glockle01a,ryckebusch97,rosner00}).
In the PWIA these are sensitive to the 
distribution of relative momenta in the nucleus. 
Consequently, we show in the following 
momentum correlations, defined as 
\begin{equation}
\label{eq:momcorr}
C^{SM_S}(\vec p) = { p^2 \over 4 \pi } \ 
< \Psi \ J=0 \ M=0 | \ \delta(\vec p_{12}-\vec p ) 
\  |S \ M_S > <S \ M_S | \  
| \Psi \ J=0 \ M=0 >  
\end{equation}
They are the probabilities to find a pair 
of nucleons in a spin state $|S \ M_S >$ and 
with a relative momentum $\vec p$. 
A similar definition in configuration space is 
given in Ref.~\cite{forest96}, 
where it has been observed numerically that this 
function has a simple angular dependence, which 
can be expanded in two Legendre polynomials 
$P_f ( \hat p \cdot \hat e_z )$ for $f=0$ and $f=2$:
\begin{equation}
\label{eq:corrangle}
C^{SM_S}(\vec p) = C^S_{f=0} (p) + C^{SM_S}_{f=2} (p) 
         \  P_2 ( \hat p \cdot \hat e_z )
\end{equation}
It only depends  on the angle 
between the momentum and the quantization axis $\hat e_z$. 
In Appendix~\ref{app:c2npw} we give an analytical 
proof of this relation. 

The probabilty for two nucleons to be in a 
fixed spin state $S$ is given by
\begin{equation}
\label{eq:spinprob}
N_{S}= \sum_{M_S} < \Psi \ J=0 \ M=0 | \ S \ M_S > <S \ M_S \  
| \Psi \ J=0 \ M=0 >  
\end{equation}
For completeness these values are given in Table~\ref{tab:nnpairprobalpha}.
In the following we will always normalize the correlations 
to $4\pi \ \int dp \ C(p)=1$. The probabilties show the 
importance of individual channels to the total correlation. 
 
\begin{table}[tp]
  \begin{center}
    \begin{tabular}{r|ll}
interaction      & $S=0$   & $S=1$  \cr
\hline
CD-Bonn          & 44.60   &  55.40 \cr
CD-Bonn+TM       & 44.98   &  55.02 \cr
AV18             & 43.07   &  56.93 \cr
AV18+TM          & 42.95   &  57.05 \cr
AV18+TM'         & 42.05   &  57.95 \cr
AV18+Urb-IX      & 41.87   &  58.13 \cr
    \end{tabular}
    \caption{Probabilities $N_{S}$ to find NN pairs in spin 
             $S=0$ and $S=1$ states
             in $^4$He as given in Eq.~(\ref{eq:spinprob}). 
             All probabilities are given in \%. }
    \label{tab:nnpairprobalpha}
  \end{center}
\end{table}

\begin{figure}
\begin{center}
\includegraphics[angle=0,width=7cm]{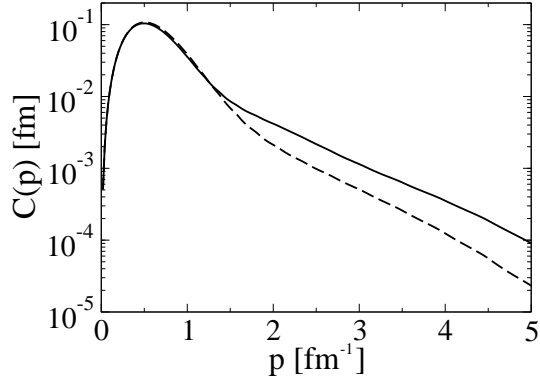}	
\end{center}

\caption{Spin-averaged NN momentum correlations in $^4$He for 
         the AV18+TM (solid line) and the CD-Bonn+TM (dashed line)
         interactions. 
         The functions are normalized to 
         $\int C(p) dp = { 1 \over 4 \pi } $.}
\label{fig:c2pnn}

\end{figure}

\begin{figure}
\begin{center}
\includegraphics[angle=0,width=7cm]{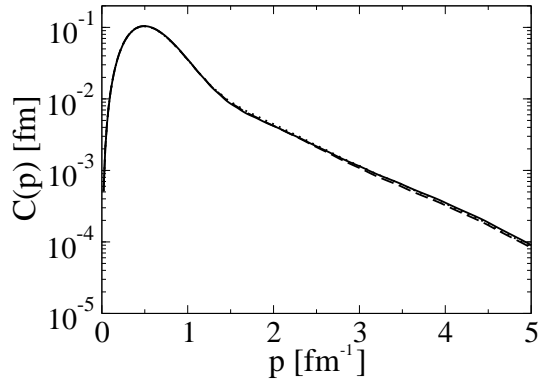}	
\end{center}

\caption{Spin-averaged NN momentum correlations in $^4$He for 
         the AV18+TM (solid line), the AV18+Urb-IX (dotted line)
         and the AV18+TM' (dashed line)
         interactions. 
         The functions are normalized to 
         $\int C(p) dp = { 1 \over 4 \pi } $.}
\label{fig:c2n3nf}

\end{figure}

In Figs.~\ref{fig:c2pnn} and \ref{fig:c2n3nf} 
spin independent momentum correlations are shown, which 
have been obtained by summing over all $SM_S$ states. 
Obviously, because of no fixed quantization axis, 
they are angular independent. 
The first figure shows the momentum correlation 
for the CD-Bonn+TM  and AV18+TM  interactions.
Similar to the distribution functions, they 
show discrepancies above $p=1$~fm$^{-1}$. In contrast, we 
did not find a similar model dependence for different 
3NF forces. This is shown in Fig.~\ref{fig:c2n3nf}
and suggests that 3NF models to not affect observables,
which are considered to be sensitive to NN correlations. 
A search for kinematical regions, where FSI's
and MEC's are suppressed, might reveal 
these correlations. In this case they  
should show up for momenta greater than  $p=1$~fm$^{-1}$.  
 
\begin{figure}
\begin{center}
\includegraphics[angle=0,width=7cm]{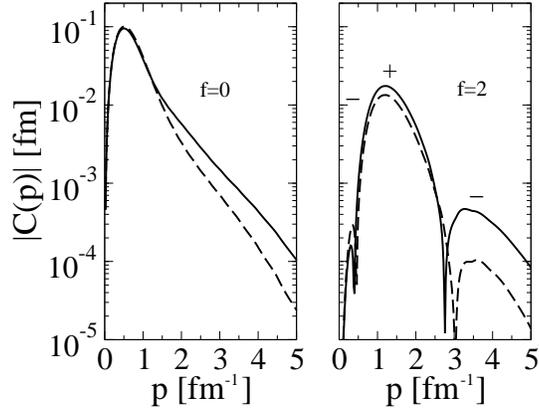}	
\end{center}

\caption{Angular independent ($f=0$) and dependent ($f=2$) parts 
         of the NN correlations $C^{S\ M_S}$
         in $^4$He for spin $S=1$ and its third component $M_S=0$, 
         as defined in the text,
         compared for different interactions on a logarithmic scale.   
         The correlation functions  are based on 
         calculations using the AV18+TM (solid lines) 
         and CD-Bonn+TM (dashed  lines) potentials.
         The functions are normalized, such that the angular 
         independent part fulfills $\int dp \ C(p)={1 \over 4\pi}$.
         The magnitude $|C|$ is shown. 
         $+$($-$) indicates positive (negative) $C_{f=2}$.}
\label{fig:c2pangnn}

\end{figure}

\begin{figure}
\begin{center}
\includegraphics[angle=0,width=7cm]{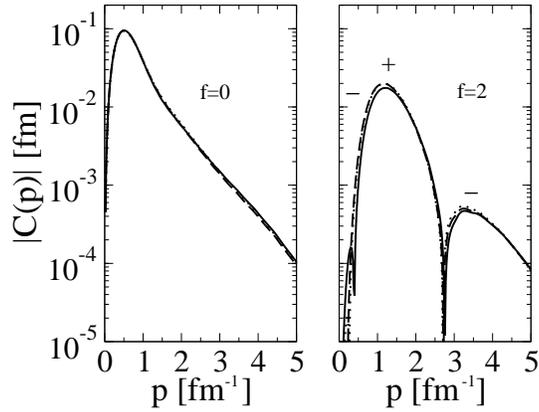}	
\end{center}

\caption{Same as Fig.~\ref{fig:c2pangnn}, except that the 
         correlation functions  are based on 
         calculations using the AV18+TM (solid lines),
         AV18+Urb-IX (dotted lines) 
         and AV18+TM' (dashed  lines) potentials.}
\label{fig:c2pang3nf}

\end{figure}

We also show the angular dependence of these momentum correlations. 
In Figs.~\ref{fig:c2pangnn} and \ref{fig:c2pang3nf} both parts 
of the correlation, as defined in Eq.~(\ref{eq:corrangle}), are 
displayed for $S=1$ and $M_S=0$. The angular 
dependent part does not depend on the 3NF, but 
for higher momenta, on the NN interaction. Around $p=1$~fm$^{-1}$ 
the $f=2$ part is  comparable in size to the $f=0$ part. 
In this region one can expect a visible angular dependence 
of the correlation. This is related to the toroidal structures 
found in configuration space correlations in \cite{forest96}.

In recent years  a knock-out 
reaction on $^4$He with $^3$H in the final state has received a great deal of
attention \cite{leeuwe98,e97-111,sofianos99,benhar00}.  
It has been shown that this reaction might be sensitive 
to the short-range correlations in nuclei \cite{morita91}. 
A first experiment has not shown the expected dip in the cross section 
\cite{leeuwe98}, which has been tracked back to effects 
of MEC's and FSI's. Ongoing experiments probe  
this reaction in different kinematical configurations, which are expected  to be 
more sensitive to the correlations. The cross sections in the PWIA  
or in the more reliable Generalized Eikonal Approximation  
approach \cite{frankfurt97} are connected 
to the $^4$He/$^3$H overlap functions 
\begin{equation}
  \label{eq:toverdef}
  T ( p ) = \sum_{m_t} 
                     < \Psi \ J=0 \ M=0  | \ \delta ( q_4 - p ) \ 
                           | \phi_t \ j_t m_t > < \phi_t \ j_t m_t | \Psi \ J=0 \ M=0 >  
\end{equation}
The momentum of the fourth  particle is fixed to $p$ and the state 
of the other three is projected on the triton state $\phi_t$ with spin $j_t={1 \over 2}$
and third component $m_t$. Because of the sum over different orientations 
of the $^3$H state, $T$ is angular independent. One can show that this is still 
true, if one fixes $m_t$ \cite{noggaphd}. 
The probability  to find a $^3$H inside 
the $\alpha$ particle is given by
$N_t = \int dp \ T(p)$. For completeness 
we give our results for $N_t$ in Table~\ref{tab:taovernorm}.
The results depend slightly on the interaction model, but 
are of the order of 80~\%. Thus, one observes a definite change in the 
3N configuration in the presence of the fourth nucleon.   

\begin{table}[tp]
  \begin{center}
    \begin{tabular}{l|r}
 interaction &  $N_t$($^4$He)    \cr
\hline   
CD-Bonn                    &  84.46        \cr
CD-Bonn +TM                &  83.49        \cr
AV18                       &  82.40        \cr
AV18+TM                    &  80.84        \cr
AV18+Urb-IX             &  80.33        \cr
AV18+TM'                   &  80.54        \cr
    \end{tabular}
    \caption{Normalization constants $N_t$ of the $^3$H-p overlap 
             distributions in $^4$He. Results are given in \%.}
    \label{tab:taovernorm}
  \end{center}
\end{table}
 
\begin{figure}
\begin{center}
\includegraphics[angle=0,width=7cm]{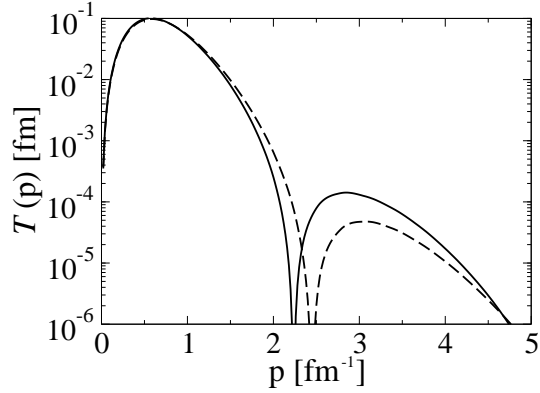}	
\end{center}

\caption{$^3$H-p momentum distribution $ T $         
         in $^4$He on a logarithmic scale.   
         The distribution functions  are based on 
         calculations using the AV18+TM (solid line) and CD-Bonn+TM (dashed line)
         potentials.
         The functions are normalized to $\int T (p) dp = { 1 \over 4 \pi } $. }
\label{fig:overnn}

\end{figure}

Fig.~\ref{fig:overnn} shows the dependence of $T$ on the NN 
interaction. The function exhibits a dip structure around 
$p=2$~fm$^{-1}$. The structure is the result of a node 
in the momentum space $s$-wave function of the fourth particle 
relative to the other three. This node is a necessary 
consequence of the short-range repulsion. Parity and 
angular momenta for $^3$H and $^4$He guarantee  
that only the $s$-wave contributes to $T$.   
The figure shows that $T$, indeed, 
depends on the NN interaction. 

\begin{figure}
\begin{center}
\includegraphics[angle=0,width=7cm]{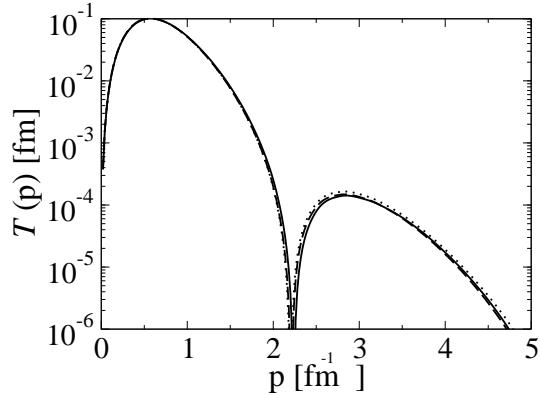}	
\end{center}

\caption{Same as Fig.~\ref{fig:overnn}, except that the distribution functions  are based on 
         calculations using the AV18+TM (solid line), AV18+Urb-IX (dotted line)  
         and AV18+TM' (dashed line) potentials.}
\label{fig:over3nf}

\end{figure}

The comparision in Fig.~\ref{fig:over3nf} of the results for different 3NF's show 
that $T$ does not depend on the 3NF's. Therefore, our results confirm 
that the measurement of $T$ might be valuable to pin down 
the correlations of two nucleons due to different NN forces
(if FSI and MEC effects would be negligible).   

\section{Conclusions and outlook}
\label{sec:concl}
We solved the Faddeev-Yakubovsky equations for the bound 
4N system in momentum space and obtained converged results. 
Two-nucleon interactions, by themselves,  underbind 
the $\alpha$ particle and leave room for considerable model dependences.
Taking properly adjusted 3NF's into account, one can considerably reduce the model 
dependences of the BE's. 
The combinations of NN and 3N forces  
lead, in general, to a small overbinding, suggesting that 
4N forces are repulsive and  much smaller than 3N forces. 

We also investigated model dependences of the 
WF. For momenta below $p=1$~fm$^{-1}$ 
we do not observe any model dependences in the 
momentum distributions and correlations. For higher momenta, 
only effects of the NN interaction show up, because the 3NF's 
do not affect these single nucleon and NN properties.

In contrast, we found a huge effect of 
3N forces on 3N correlations
visible in the matrix elements of the 3N force.
These effects require further visualization in future studies.
We also found that that the $\alpha$ particle ground state
is an extremely pure $T=0$ isospin state. The admixtures of 
$T=1$ and $T=2$ states are of the order of 0.003~\% and 0.005~\%,
respectively. This sharpens and questions 
the result found before in \cite{ramavataram94}.

These calculations provide a baseline for 
the analysis of experiments involving the $\alpha$ particle, which 
require highly accurate WF's and insight 
into NN-force model dependences. The technical developments 
presented are 
also important for further studies of nuclear 
interactions based on $\chi$PT. 
First studies have already been started \cite{epelbaum01a,epelbaum01c,epelbaum01d}.
$\chi$PT allows a systematic 
derivation of 3NF's, which are consistent with the NN forces. 
An investigation of these 3N forces requires accurate 
techniques for solving the 3N and 4N Schr\"odinger 
equation, in order to fix the parameters of the force 
and to see their effects. The bound states are an intersting 
object for these studies, because they are the physical quantities 
very sensitive to 3NF effects and are dominated by the low-energy 
properties of the nuclear interaction.   
    
\section*{Acknowledgments}

A.N. and B.R.B. acknowledge partial support from 
NSF grant\# PHY0070858. The numerical calculations 
have been performed on the Cray T3E of the NIC in J\"ulich,
Germany.

\appendix 

\section{Partial wave decomposition of correlation functions}

\label{app:c2npw}

The spin-dependent correlation functions are angular dependent. 
In momentum space and for a general nuclear $A$-body
bound-state $\Psi$ with angular momentum $J\ M$, 
it is defined as 
\begin{equation}
\label{eq:momcorrb}
C^{SM_S}(\vec p) = { p^2 \over 4 \pi } \sum_M \ 
< \Psi \ J\ M| \ \delta(\vec p_{12}-\vec p ) 
\  |S \ M_S > <S \ M_S | \  
| \Psi \ J\ M >  
\end{equation}
The operator $\delta(\vec p_{12}-\vec p ) 
\  |S \ M_S > <S \ M_S |$ acts only on the subsystem
of particles 1 and 2, i.e., (12). 
Therefore we choose coordinates, which single out this 
subsystem and denote the coordinates of the remaining 
particles by $\alpha_{A-2}\ J_{A-2}\ M_{A-2}$, where  we have separated 
the angular momentum quantum numbers. $\alpha_{A-2}$ also includes 
the motion of the (12) subsystem relative to the $A-2$ 
spectators. The two-body subsystem is described by the usual 
momentum $p_{12}$ and quantum numbers $l_{12}$, $s_{12}$ and $j_{12}$
and the third component $m_j$.
Resolving the coupling of the angular momemtum of the (12) 
subsystem and the spectators to the total angular momentum,
one obtains for the correlation 
\begin{eqnarray}
  \label{eq:defcorra2}
  C^{S \ M_S}( \vec p )  & = &
{ 1 \over 2 J +1 } \ \sum_{M}  
    \sum_{\begin{array}{c} \alpha_{A-2} \\ J_{A-2} M_{A-2} \end{array}}
\sum_{\begin{array}{c} l_{12}l_{12}'s_{12}s_{12}' \\ j_{12}j'_{12} m_j m_j ' \end{array}} 
\int dp_{12} \ p_{12}^2 \ \int dp'_{12} \ {p'_{12}} ^2  \nonumber \\[5pt] 
& & \quad ( j_{12} J_{A-2} J, m_j M_{A-2} M ) \ ( j'_{12} J_{A-2} J, m'_j  M_{A-2}  M )  
               \nonumber \\[5pt]
& & \quad  < \Psi \ JM | p_{12} \ \alpha_{A-2} \ ((l_{12}s_{12})j_{12}  J_{A-2}) JM > 
           < p'_{12} \ \alpha_{A-2} \ ((l'_{12}s'_{12})j'_{12}  J_{A-2}) JM 
                | \Psi \ JM > \nonumber \\[5pt] 
& & \quad < p_{12} (l_{12} s_{12}) j_{12} m_j | 
               \ \delta^{3} (\vec p - \vec p_{12}) \ |S\ M_S >< S\ M_S|  
         \  | p'_{12} (l'_{12} s'_{12}) j'_{12} m'_j >
\end{eqnarray}
The nuclear bound state WF 
$< p'_{12} \ \alpha_{A-2} \ ((l'_{12}s'_{12})j'_{12}  J_{A-2}) JM   
              | \Psi \ JM >$ is independent of $M$. 
We choose $M=J$ in these matrix elements and perform the $M$ and 
$M_{A-2}$ summation, using the orthogonality relations for the 
Clebsch-Gordan coefficients. This leads to 
\begin{eqnarray}
  \label{eq:defcorra3}
  C^{S \ M_S}( \vec p )  & = &
\sum_{ \alpha_{A-2} J_{A-2}}
\sum_{\begin{array}{c} l_{12}s_{12} j_{12} \\ l_{12}' s_{12}'  \end{array} } 
\int dp_{12} \ p_{12}^2 \ \int dp'_{12} \ {p'_{12}} ^2 \nonumber \\[5pt]  
& & \quad \ { 1 \over 2 j_{12} +1 } \sum_{m_j}   
< p_{12} (l_{12} s_{12}) j_{12} m_j | 
               \ \delta^{3} (\vec p - \vec p_{12}) \ |S\ M_S >< S\ M_S|  
         \  | p'_{12} (l'_{12} s'_{12}) j_{12} m_j >
 \nonumber \\[5pt] 
& & \quad  < \Psi \ JJ | p_{12} \ \alpha_{A-2} \ ((l_{12}s_{12})j_{12}  J_{A-2}) JJ > 
           < p'_{12} \ \alpha_{A-2} \ ((l'_{12}s'_{12})j_{12}  J_{A-2}) JJ
                | \Psi \ JJ >     
\end{eqnarray}
In this form the problem is reduced for arbitrary nuclei 
to the matrix element 
\begin{equation}
M_{12} \equiv
{ 1 \over 2 j_{12} +1 } \sum_{m_j}   
< p_{12} (l_{12} s_{12}) j_{12} m_j | 
               \ \delta^{3} (\vec p - \vec p_{12}) \ |S\ M_S >< S\ M_S|  
         \  | p'_{12} (l'_{12} s'_{12}) j_{12} m_j > ,
\end{equation}
which is diagonal in $j_{12}$ and $m_{j}$. 

By inserting the unity operator in states of 3D momentum and resolving 
the coupling of spins and orbital angular momenta, we are able to simplify 
the expression to 
\begin{eqnarray}
  \label{eq:part2ncorrb}
  M_{12} & = &   \delta_{s_{12} s'_{12}} \delta_{s_{12} S}  
    { \delta (p_{12}-p) \over p_{12} p}  \ { \delta (p'_{12}-p) \over p'_{12} p}  \  
           { 1 \over 2 j_{12} +1 } \sum_{m_j}  \nonumber \\[5pt]
& & \quad (l_{12} S j_{12} , m_j-M_S \ M_S ) 
           \ (l'_{12} S j_{12} , m_j-M_S \ M_S )   \ 
    Y^*_{l_{12}m_j-M_S}(\hat p) \ Y_{l'_{12}m_j-M_S}(\hat p) 
\end{eqnarray}
Using standard techniques, one can recouple the angular momenta 
to obtain a coupled spherical harmonic ${\cal Y}_{l_{12} l'_{12}}^{f\mu}(\hat p \hat p ) $. It turns out that only $\mu=0$ contributes,
which is expected, because fixing 
the spin only fixes the $z$-axis. The matrix elements 
depend only on $x=\hat p \cdot \hat e_z$. This dependence can be 
expanded in Legendre polynomials and one ends up with
\begin{eqnarray}
  \label{eq:part2ncorre}
  M_{12} & = &  \delta_{s_{12} s'_{12}} \delta_{s_{12} S}  
    { \delta (p_{12}-p) \over p_{12} p}  \ { \delta (p'_{12}-p) \over p'_{12} p}  \  
           \sum_f (-)^{S-j_{12}} (-)^{l_{12}+l'_{12}-f} \ 
            \sqrt{ (2 l_{12} +1 ) (2 l'_{12} +1 ) (2 f + 1 )\over 2 S + 1 } \ 
      \nonumber \\[5pt]
& & \quad  \left\{      
      \begin{array}{ccc} 
        S & l_{12}  & j_{12} \\
        l'_{12} & S &  f      
      \end{array} \right\} 
          (S \ f \ S , M_S \ 0 )  \ (l_{12} l'_{12} f,00) \ { 1 \over 4 \pi } \ P_f (x)
\end{eqnarray}
Because $S$ is restricted to 0 and 1, the order of the Legendre polynomial $f$
can only take the values 0, 1 and 2. Parity conservation fixes 
the phase $(-)^{l_{12}+l'_{12}}=1$. Therefore, the Clebsch-Gordan coefficient 
$(l_{12}l'_{12}f,00)$ demands even $f$'s. Because of this, 
the expansion of the angular dependence contains  only two 
Legendre polynomials: $P_0(x)$ and $P_2(x)$. This proves 
the form of Eq.~(\ref{eq:corrangle}). From the explicit form 
of $M_{12}$ one also reads off that the $M_S$ dependence 
is given by an overall  Clebsch-Gordan coefficient. 
This justifies the fact that we only present results for $M_S=0$
in Section~\ref{sec:resc}.
We also see that the $f=0$ part of $C$  
is independent of $M_S$. 
Finally, we would like to note 
that the expressions are also valid in configuration space,
replacing the momenta by the corresponding distances.

\section{Treatment of the 3NF embedded in the 4N Hilbert space}
\label{app:3nf}

\subsection{TM-like forces}

We consider the 3NF as the successive applications of NN-like 
interactions, which, however, do not respect parity and 
rotational invariance. Only the full 3NF 
repects these symmetries \cite{huber97b}. The 
YE's~\ref{eq:alphaeq1} and \ref{eq:alphaeq4} require the matrix 
elements 
\begin{equation}
  \label{eq:3nfmat}
   < (12)3,4 | V_{123}^{(3)} | \Psi > 
\end{equation}
where we can assume that the state $\Psi$  is antisymmetric 
in the nucleons 123. 

One distinguishes four terms in the TM force, the so called $a$-, $b$-, $c$- and $d$-term,
which are given by their individual strength constants.
These constants are listed together with $V_0$ in 
Table~\ref{tab:3nfpara}. 
\begin{eqnarray}
  \label{eq:tmform}
& &   V_{123}^{(3)} = V_0 \ \big( \ 
     a \ \vec {\tau_1} \cdot \vec {\tau_2} \ W^a_{23} \ \ W^a_{31} 
    +b \ \vec {\tau_1} 
          \cdot \vec {\tau_2} \ \vec W^b_{23} \ 
                              \cdot  \vec W^b_{31} 
    +c \ \vec {\tau_1} \cdot \vec {\tau_2}  
           \ (W^a_{23} \ \ W^c_{31} + W^c_{23} \ W^a_{31}) 
    +d \ \vec {\tau_3} \cdot \vec {\tau_1} 
             \times  \vec {\tau_2} \ \vec W^d_{23} \ 
                              \cdot  \vec W^b_{31} \big) \nonumber \\[5pt]
& & 
\end{eqnarray}
where we have separated the isospin operators 
(Pauli isospin matrices $\tau_i$) and the spin-orbital operators $W$. 

\begin{table} 
  \begin{center}
    \begin{tabular}{l|r|rrrr}
  &    $4 (2\pi)^6 \ V_0$ [$m_N^{-2}$]  
       & a [$m_\pi^{-1}$] & b [$m_\pi^{-3}$] & c [$m_\pi^{-3}$] & d [$m_\pi^{-3}$] \cr
\hline
TM    &  179.7                       
       & 1.13             & -2.58            & 1.00              & -0.753 \cr
TM'   &  179.7                       
      & -0.87             & -2.58            & 0.00              & -0.753 
    \end{tabular}
    \caption{Strength constants of the TM \cite{coon79} and TM' \cite{huber97b}
             3NF's.
             The numbers are in units of the nucleon mass $m_N=938.926$~MeV 
             and the $\pi$ mass $m_\pi=139.6$~MeV.}
    \label{tab:3nfpara}
  \end{center}
\end{table}

\begin{figure}[tp]
\begin{center}
\includegraphics[angle=0,width=7cm]{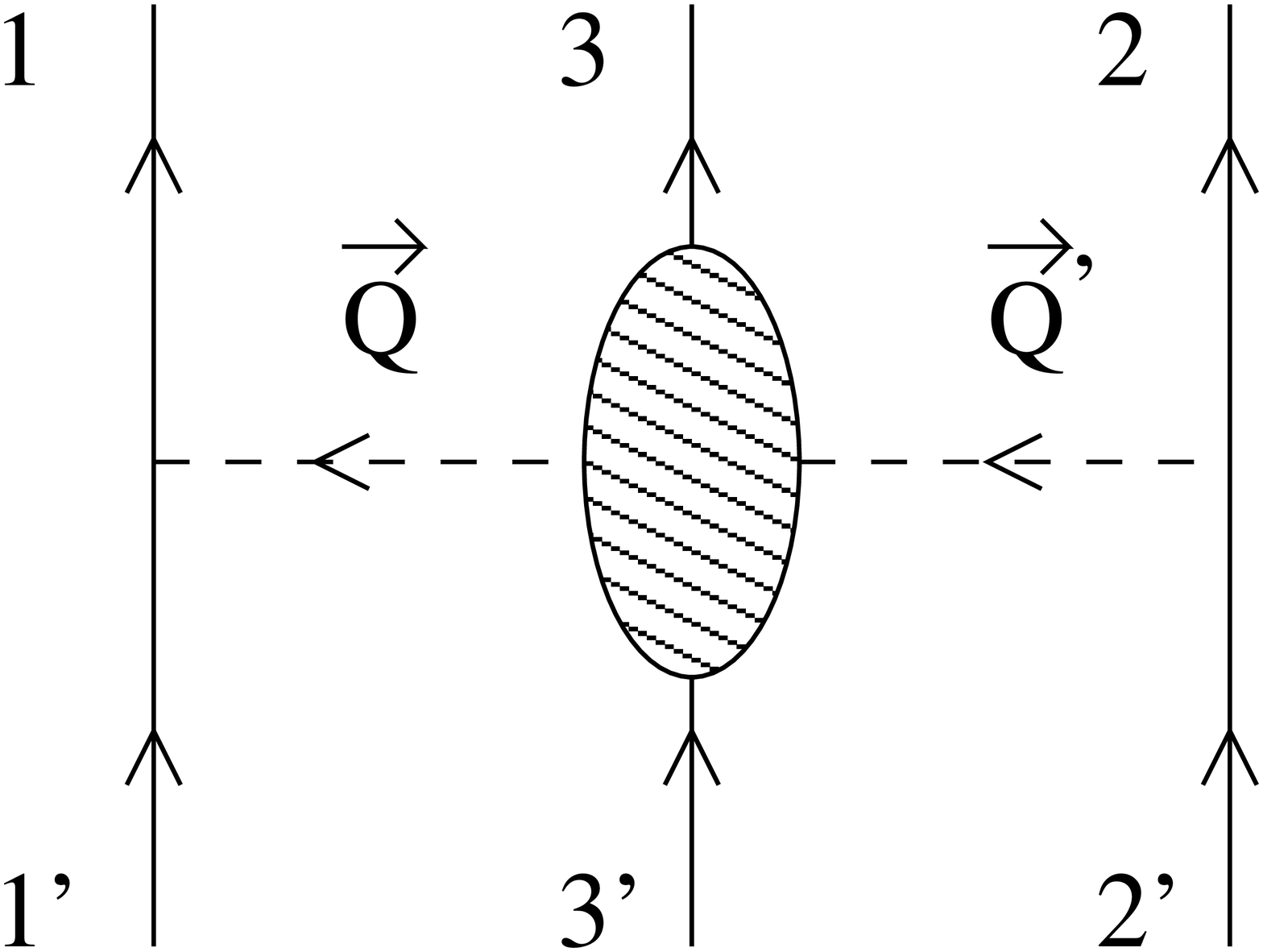}	
\end{center}
\caption{\label{fig:3nfqmom} Symbolic representation of a 3NF, like the TM force, 
         and the definition of the 
         momentum transfers $\vec Q$ and $\vec Q^{\ \prime}$ within the two subsystems }
\end{figure}

The $W$'s can be read off from the definition of the TM force in momentum space, 
as given in Ref.~\cite{huber97b}. 
\begin{equation}
  \label{eq:tmws}
\begin{array}{ll}
\displaystyle
W_{23}^a = F({\vec {Q'}}^2) { \vec \sigma _2 \cdot \vec {Q'}  \over {\vec {Q'}}^2 + m_\pi^2 } &
\displaystyle
W_{31}^a = F({\vec Q}^2) { \vec \sigma _1 \cdot \vec Q  \over {\vec Q}^2 + m_\pi^2 } \cr
\displaystyle
\vec W_{23}^b = F({\vec {Q'}}^2) { \vec \sigma _2 \cdot \vec {Q'}  
                 \over {\vec {Q'}}^2 + m_\pi^2 } \vec {Q'} &
\displaystyle
\vec W_{31}^b = F({\vec Q}^2) { \vec \sigma _1 \cdot \vec Q  
                 \over {\vec Q}^2 + m_\pi^2 } \vec Q \cr
\displaystyle   
W_{23}^c =  F({\vec {Q'}}^2) { \vec \sigma _2 \cdot \vec {Q'}  
                \over {\vec {Q'}}^2 + m_\pi^2 } {\vec {Q'}}^2 & 
\displaystyle      
W_{31}^c = F({\vec Q}^2) { \vec \sigma _1 \cdot \vec Q  \over {\vec Q}^2 + m_\pi^2 } {\vec Q}^2 \cr 
\displaystyle
\vec W_{23}^d =  F({\vec {Q'}}^2) { \vec \sigma _2 \cdot \vec {Q'}  
                          \over {\vec {Q'}}^2 + m_\pi^2 } 
                        \vec \sigma_3 \times \vec {Q'} & \cr
\end{array}
\end{equation}
with the momentum transfers $  \vec Q   = \vec k_1 - \vec k_1 ^{\ \prime} $ and
$ \vec Q^{\ \prime} =  \vec k_2^{\ \prime} - \vec k_2 $, as indicated in Fig.~\ref{fig:3nfqmom}.
The $\sigma_i$'s are Pauli spin matrices and the form factors are chosen 
to be 
$ F({\vec Q}^2) = { \Lambda^2 - m_\pi^2 \over \Lambda^2 + {\vec Q}^2} $.

Applied to a state 
vector $\psi$, all four terms  have the form 
\begin{equation}
  \label{eq:tmformb}
  \psi ' \sim W_{23} \ I \ W_{31} \ \Psi ,
\end{equation}
where we have abbreviated the isospin operators by $I$.

By introducing the unit operator in the coordinates, which are natural for the $W$ potentials,
we are able to turn Eq.~(\ref{eq:tmformb}) into 
\begin{eqnarray}
  \label{eq:tmmatform}
  < (12)3,4 | \psi ' > &\sim & < (12)3,4 | (23)1,4 ' > < (23)1,4 ' | W_{23} | (23)1,4 ''>  \cr
& &       <(23)1,4 '' | I | (31)2,4 ''' > 
       < (31)2,4 ''' | W_{31} | (31)2,4^* > < (31)2,4^* | \Psi >  .
\end{eqnarray}
We omit the integrals and sums over momenta and quantum numbers of the 
intermediate states, in order  to simplify the expressions and denote by 
$(ij)k,l$ Jacobi coordinates, which single out the pair $ij$, the 3-body 
cluster $ijk$ and the spectator $l$.
$\Psi$ originally enters in $(12)3,4$ coordinates. But because of the antisymmetry of $\Psi$
in the $(123)$ subsystem, 
the $(31)2,4$ coordinates are equivalent in this case. 

Because the $W_{23}$'s do not respect the symmetries of nuclear interactions,
the sum over $''$- and $'''$-states have to include other 
parities or other total angular momenta, depending on whether 
$a/c$ of $b/d$-terms are considered. The $'$-, $''$- and $'''$-sums 
have also to include unphysical symmetric states of the (31) or (23) 
subsystems. The matrix elements of the coordinate transformations 
 $ <(23)1,4 '' | I | (31)2,4 ''' > $ are given in Ref.~\cite{nogga01b,noggaphd}. 
The isospin operator leads to a change of the isospin part of the transformation. 
The new isospin matrix elements have been derived in Ref.~\cite{huber97b} 
for the 3N system and are given below for the 4N system for completeness.
 
The matrix elements for the different $W$'s  are summarized  
below.
For the $a$-term one finds
\begin{eqnarray}
  \label{eq:waterm}
& &   <(31)2,4 | W_{31}^a | (31)2,4 ' >  \nonumber \\[5pt]
   & & \qquad ={ \delta(q_4 - q_4 ' ) \over q_4 q_4' } \ { \delta(p_2 - p_2 ' ) \over p_2 p_2' } \
         \delta_{I_4 I_4'} \delta_{I_2 I_2'} \delta_{l_4 l_4'} \delta_{l_2 l_2'} 
         \delta_{J J'} \delta_{M M'}\delta_{j_3 j_3'} 
          \delta_{j_{31}j_{31}'}\delta_{|l_{31}-l_{31}'|,1} \cr
&& \quad \qquad           \ 2 \pi \sqrt{6} \ (-)^{j_{31}+1+{\rm max}(l_{31},l_{31}')} 
         \sqrt{\hat s_{31} \hat s_{31}'} 
          \left\{ 
            \begin{array}{ccc}
               { 1 \over 2 } & { 1 \over 2 } & s_{31}' \cr
                    1        &    s_{31}     &  { 1 \over 2 } \cr 
            \end{array} 
          \right\} 
          \left\{ 
            \begin{array}{ccc}
              l_{31}' & s_{31}' & j_{31}' \cr
              s_{31}  & l_{31}  &  1       \cr 
            \end{array} 
          \right\}   \cr
& & \quad \qquad \sqrt{{\rm max}(l_{31},l_{31}')} 
             \ \left( p_{31} \ H_{l_{31}'}(p_{31},p_{31}')
                     -p_{31}' \ H_{l_{31}}(p_{31},p_{31}') \right)  
\end{eqnarray}
The operator has no isospin dependence; therefore, it is diagonal in
isospin space.  The momentum dependence is given in terms 
of the function $H$, which is a combination of Legendre polynomials
of the second kind $Q_l$ and their derivatives $Q'_l$.
\begin{equation}
  \label{eq:hfunc}
  H_l (p,p') = { 1 \over p p' } \ ( Q_l(B_{m_\pi}) - Q_l(B_{\Lambda}) ) 
                  + { \Lambda^2 - m_{\pi}^2  \over 2 ( p p' )^2 } \ Q_l' (B_\Lambda)   
\end{equation}
with $B_{m_\pi} = { p^2 + {p'}\; ^2 + m_\pi^2  \over 2 p p ' } $
and $ B_{\Lambda} = { p^2 + {p'}\; ^2 + \Lambda ^2  \over 2 p p ' } $.

The $c$-term looks very similar as the $a$-term and follows, if one replaces $H$ 
in Eq.~(\ref{eq:waterm}) by 
\begin{equation}
  \label{eq:htildefunc}
  \tilde H_l (p,p') = - { m_\pi^2 \over p p' } \ ( Q_l(B_{m_\pi}) - Q_l(B_{\Lambda}) ) 
                  - { \Lambda^2 - m_{\pi}^2  \over 2 ( p p' )^2 } \ \Lambda^2 
                      \ Q_l' (B_\Lambda)   .  
\end{equation}
For our convenience, we used the abbreviation $\hat k = 2k +1$ in these expressions. 
The notation of the different quantum numbers is an obvious generalization 
of the notation in Fig.~\ref{fig:jacobi}.

Because the $W_{23}$- and $W_{31}$-operators are equivalent up to a renumbering 
of the particles, the matrix elements are equal up to a phase factor
\begin{eqnarray}
  \label{eq:wacphase}
  <(23)1,4 | W^{a,c}_{23} | (23)1,4 '> 
     & = & (-)^{l_{31}+s_{31}+t_{31}+l_{31}'+s_{31}'+t_{31}'+1} 
             \  <(31)2,4 | W^{a,c}_{31} | (31)2,4 '> .
\end{eqnarray}

As we already mentioned, we replace the simple 
isospin transformation matrix element by a combination of the transformation 
and the isospin operator
\begin{equation}
  \label{eq:isotau12}
< (23)1,4 | \vec \tau_1 \cdot  \vec \tau_2 | (31)2,4 '> 
   = \delta_{TT'} \delta_{M_T M_T'} \delta_{\tau \tau'}
    (-6) \ (-)^{t_{23}} \sqrt{ \hat t_{23} \hat t_{31}'} 
       \ \left\{ 
            \begin{array}{ccc}
              {1 \over 2 } & {1 \over 2 } &   t_{31}'      \cr
              {1 \over 2 } &      1       & {1 \over 2 }  \cr
                t_{23}     & {1 \over 2 } &   \tau         \cr
            \end{array}
          \right\}    .
\end{equation}

The $b$- and $d$-terms are a slightly more complicated, because 
the NN-like potentials are now vector operators. 
The matrix elements of the spherical components 
   $W^{1} =   - { 1 \over \sqrt{2} } \ ( W^x+iW^y )$,  
$W^{0} =   W^z$ and $W^{-1} =   { 1 \over \sqrt{2} } \ ( W^x-iW^y )$  
decompose into a Clebsch-Gordan coefficient 
and a reduced matrix element
\begin{equation}
  \label{eq:reduced}
  <(31)2,4 | W^\mu_{31}|(31)2,4 '>   =  (J'1J,M' \mu M ) \ <(31)2,4 || W_{31}||(31)2,4 '> .
\end{equation}
The scalar product in spherical coordinates reads 
\begin{equation}
  \label{eq:skalarsphaerical}
   \vec W_{23} \cdot \vec W_{31} = \sum_{\mu} (-)^\mu  W_{23}^\mu W_{31}^{-\mu}.
\end{equation}

Because there is no dependence on the third component of the total 
angular momentum, neither in the transformation matrix elements nor  
in the incoming state, we can analytically perform the $M''$ and $\mu$-summations
\begin{equation}
  \label{eq:mmusum}
\sum_{M'' \mu} (-)^\mu (J'' 1 J',M'' \mu M' ) \ (J^* 1 J '' , M^* -\mu M'') 
        = \delta_{J'J^*} \delta_{M'M^*} \sqrt { \hat J '' \over \hat J' } (-)^{J''-J'}
\end{equation}
and recover the conservation of the total angular momentum. 
The NN-like potentials effectively 
requires only the application of the reduced matrix elements
and the additional factor  $\sqrt { \hat J '' \over \hat J' } (-)^{J''-J'}$.
The  intermediate states are also $M$ independent. 
 
The generalization of the formulas given in Ref.~\cite{huber97b} to 
the four-nucleon system yields
\begin{eqnarray}
  \label{eq:btermredmat}
& & < (31)2,4  || W_{31}^b || (31)2,4  >' \nonumber \\[5pt]
& & \quad = { \delta(q_4 - q_4 ' ) \over q_4 q_4' } 
                                            \ { \delta(p_2 - p_2 ' ) \over p_2 p_2' } \
                   \delta_{I_4 I_4'} \delta_{I_2 I_2'} \delta_{l_4 l_4'} \delta_{l_2 l_2'} 
(-)^{J'+j_3'+j_3+I_4+I_2+s_{31}+s_{31}'} 
\sqrt{\hat J' \hat j_3' \hat j_3 \hat s_{31}' \hat s_{31} \hat j_{31}' \hat j_{31}} \cr     
&  & \qquad  \qquad 
\left\{
 \begin{array}{ccc}
 j_3'   &  1   & j_3 \cr
 J      &  I_4 & J'  \cr
 \end{array}
\right\} \
\left\{
 \begin{array}{ccc}
 1   &  j_{31}   & j_{31}' \cr
 I_2 &  j_3'     & j_3     \cr
 \end{array}
\right\}\
\left\{
 \begin{array}{ccc}
 {1 \over 2}   &  {1 \over 2}   & s_{31} \cr
 1             &  s_{31}'       & {1 \over 2}    \cr
 \end{array}
\right\} \cr
& & \qquad \times \left[ \delta_{l_{31}l_{31}'} \ { 2 \pi \over 3 } \ \sqrt{6} \ (-)^{l_{31}+1} \
\left\{
 \begin{array}{ccc}
 j_{31}'   & j_{31}   & 1 \cr
 s_{31}    & s_{31}'  & l_{31}\cr
 \end{array}
\right\} \        \tilde H_{l_{31}} (p_{31} p_{31}') \right. \nonumber \\[5pt]
& & \qquad \quad 
- 40 \pi \sqrt{6} \ (-)^{s_{31}'+j_{31}} \ 
\left\{
 \begin{array}{ccc}
 2         &     1     & 1 \cr
 l_{31}    & s_{31}    & j_{31} \cr
 l_{31}'   & s_{31}'   & j_{31}' \cr
 \end{array}
\right\} \   \sum_{\bar l } \hat {\bar l} H_{\bar l} (p_{31} p_{31}') \cr
& & \left. \qquad \qquad 
\times \sum_{a+b=2} { {p_{31}'} \! \! ^a {p_{31}} \! \! ^b \over \sqrt{ (2a)! (2b)!}} \
\left\{ 
\begin{array}{ccc}
 b         & a        & 2 \cr
 l_{31}'   & l_{31}   & \bar l\cr
 \end{array} \right\}
 (a \ \bar l \ l_{31}',00) \  (b \ \bar l \ l_{31},00) \right]
\end{eqnarray}
and 
\begin{eqnarray}
  \label{eq:dtermredmat}
& & < (23)1,4  || W_{23}^d || (23)1,4  >' \nonumber \\[5pt]
& & \quad = { \delta(q_4 - q_4 ' ) \over q_4 q_4' } 
                                            \ { \delta(p_1 - p_1 ' ) \over p_1 p_1' } \
                   \delta_{I_4 I_4'} \delta_{I_1 I_1'} \delta_{l_4 l_4'} \delta_{l_1 l_1'} 
(-)^{J'+j_3'+j_3+I_4+I_1+s_{23}+s_{23}'+1} 
\sqrt{\hat J' \hat j_3' \hat j_3 \hat s_{23}' \hat s_{23} \hat j_{23}' \hat j_{23}} \cr     
&  & \qquad  \qquad 
\left\{
 \begin{array}{ccc}
 j_3'   &  1   & j_3 \cr
 J      &  I_4 & J'  \cr
 \end{array}
\right\} \
\left\{
 \begin{array}{ccc}
 1   &  j_{23}   & j_{23}' \cr
 I_1 &  j_3'     & j_3     \cr
 \end{array}
\right\} \cr
& & \qquad \times \left[ \delta_{l_{23}l_{23}'} \ i \ 4\pi \ \sqrt{6} \ (-)^{l_{23}+s_{23}} \
\left\{
 \begin{array}{ccc}
 l_{23}    & s_{23}   & j_{23} \cr
 1         & j_{23}'  & s_{23}'\cr
 \end{array}
\right\} \  
\left\{
 \begin{array}{ccc}
 1         & 1           & 1 \cr
{1\over 2} & {1\over 2}  & s_{23}' \cr 
{1\over 2} & {1\over 2}  & s_{23}  \cr 
 \end{array}
\right\} \        \tilde H_{l_{23}} (p_{23} p_{23}') \right. \nonumber \\[5pt]
& & \qquad \quad 
+ \ i \ 240 \pi \sqrt{6} \ (-)^{j_{23}'} \
\sum_{\chi} (-)^{\chi} \ \hat \chi  \ 
\left\{
 \begin{array}{ccc}
 2         &  \chi     & 1 \cr
 1         &     1     & 1      \cr
 \end{array}
\right\} \  
\left\{
 \begin{array}{ccc}
 2         &  \chi     & 1 \cr
 l_{23}'   &  s_{23}'  & j_{23}'\cr
 l_{23}    &  s_{23}   & j_{23} \cr
 \end{array}
\right\} \ 
\left\{
 \begin{array}{ccc}
 1          &     1     & \chi   \cr
 {1 \over 2}&{1 \over 2}& s_{23}'\cr
 {1 \over 2}&{1 \over 2}& s_{23} \cr
 \end{array}
\right\} \   \cr
& & \left. \qquad \qquad 
\times \sum_{\bar l } \hat {\bar l} H_{\bar l} (p_{23} p_{23}')  \ 
\sum_{a+b=2} { {p_{23}'} \! \! ^a {p_{23}} \! \! ^b \over \sqrt{ (2a)! (2b)!}} \
\left\{ 
\begin{array}{ccc}
 b         & a        & 2 \cr
 l_{23}'   & l_{23}   & \bar l\cr
 \end{array} \right\}
 (a \ \bar l \ l_{23}',00) \  (b \ \bar l \ l_{23},00) \right].
\end{eqnarray}
The momentum dependent functions $H$ and $\tilde H$ are given in 
Eqs.~(\ref{eq:hfunc}) and (\ref{eq:htildefunc}). 

Again, there is a simple phase relation 
between $W^b_{23}$ and $W^b_{31}$
\begin{eqnarray}
  \label{eq:wbdphase}
  <(23)1,4 | W^{b,d}_{23} | (23)1,4 '> 
        & = & (-)^{l_{31}+s_{31}+t_{31}+l_{31}'+s_{31}'+t_{31}'} 
               \  <(31)2,4 | W^{b,d}_{31} | (31)2,4 '> 
\end{eqnarray}

The isospin matrix element of the $d$-term differs from the 
one for the $a$-, $b$- and $c$-term, given in Eq.~(\ref{eq:isotau12}).
It reads 
\begin{eqnarray}
  \label{eq:isotaucross}
< (23)1,4 | \vec \tau_3 \cdot  ( \vec \tau_1 \times \tau_3) | (31)2,4 '> & = & \delta_{TT'} \delta_{M_T M_T'} \delta_{\tau \tau'}
    24i  \ (-)^{2 \tau } \sqrt{ \hat t_{23} \hat t_{31}'} \cr
& &        \sum_{\lambda} (-)^{3 \lambda + { 1 \over 2 }} \   
       \ \left\{ 
            \begin{array}{ccc}
              \lambda      & {1 \over 2 } &   1            \cr
              {1 \over 2 } & {1 \over 2 } &  t_{23}        \cr
            \end{array}
          \right\}    
       \ \left\{ 
            \begin{array}{ccc}
              \tau         & {1 \over 2 } &   t_{23}      \cr
              {1 \over 2 } &      1       & \lambda       \cr
                t_{31}'    & {1 \over 2 } &  {1 \over 2 } \cr
            \end{array}
          \right\}.    
\end{eqnarray}

\subsection{Urbana type  forces}

The functional form of the Urbana 3NF is much simpler.
One usually expresses the Urbana interaction in terms of  
commutator and anticommutator parts. This reads
\begin{eqnarray}
  \label{eq:vurbcomm}
  V_{123}^{(3)} & = & \phantom{ + } A_{2\pi} \ [ \  \{ X_{23},X_{31} \} \  
                                 \{ \vec \tau _2 \cdot \vec \tau _3 , 
                                    \vec \tau _3 \cdot \vec \tau _1 \}  \cr 
& & \phantom{ + A_{2\pi} } + { 1 \over 4 } \   [  X_{23},X_{31} ] \  
                                 [ \vec \tau _2 \cdot \vec \tau _3 , 
                                    \vec \tau _3 \cdot \vec \tau _1 ] \ ] \cr
& & + U_0 \ T_\pi ^2 (r_{23}) \ T_\pi ^2 (r_{31}) 
\end{eqnarray}

The force is explicitly defined in terms of NN interactions 
\begin{equation}
  \label{eq:xterms}
  X_{ij} = Y_\pi ( r_{ij} ) \ \vec \sigma_i \cdot \vec \sigma_j 
              +  T_\pi (r_{ij}) \ S_{ij}
\end{equation}
$X_{ij}$ is derived from the $\pi$ exchange NN force. 
Therefore, it has a spin-spin part $\vec \sigma_i \cdot \vec \sigma_j$ 
and a tensor part 
\begin{equation}
  \label{eq:tensorop}
  S_{ij} = 3 \ \vec \sigma_i \cdot \hat r_i \ \vec \sigma_j \cdot \hat r_j 
- \vec \sigma_i \cdot \vec \sigma_j .
\end{equation}
The radial dependence is given as 
\begin{eqnarray}
  \label{eq:radialfunc}
   Y_\pi (r)  & = & { e ^ { - m_\pi r } \over m_\pi  r } \ \left( 1 - e ^ { - c r^2 } \right) \cr
   T_\pi (r)  & = & \left[ 1 + { 3 \over  m_\pi r } + { 3 \over  (m_\pi r)^2 } \right] 
        { e ^ { - m_\pi r } \over m_\pi  r } \ \left( 1 - e ^ { - c r^2 } \right)^2 
\end{eqnarray}
The parameters $A_{2\pi}$, $U_0$ and $c$ for the Urb-IX   
are given in Ref.~\cite{pudliner97}.

In \cite{witala01a} it is shown that the application of the Urbana 
force can be rewritten as 
\begin{eqnarray}
  \label{eq:urballterms}
  < (12)3,4 | \psi ' > &= & 2 A_{2\pi} \ 
  < (12)3,4 | (23)1,4 ' > < (23)1,4 ' | X_{23} | (23)1,4 ''>  \cr
& &   \quad     <(23)1,4 '' | I^- | (31)2,4 ''' > 
                  < (31)2,4 ''' | X_{31} | (31)2,4^* > < (31)2,4^* | \psi >  \cr
& &     + U_0 \ < (12)3,4 | (23)1,4 ' > < (23)1,4 ' | T^2_\pi(r_{23}) | (23)1,4 ''>  \cr
& &       <(23)1,4 '' | (31)2,4 ''' > 
       < (31)2,4 ''' | T^2_\pi(r_{31}) | (31)2,4^* > < (31)2,4^* | \psi > .
\end{eqnarray}
The isospin operators are very similar to the ones encountered 
in the TM force
\begin{eqnarray}
  \label{eq:isoplusminus}
  I^- & \equiv &  2 ( \vec \tau _1 \cdot \vec \tau _2 
      - { i \over 4 } \ \vec \tau_3 \cdot  \vec \tau _1 \times  \vec \tau _2 )\cr
  I^+ & \equiv &  2 ( \vec \tau _1 \cdot \vec \tau _2 
      + { i \over 4 } \ \vec \tau_3 \cdot  \vec \tau _1 \times  \vec \tau _2 ) .
\end{eqnarray}
It is easy to combine Eqs.~(\ref{eq:isotau12}) and (\ref{eq:isotaucross}) to 
find their matrix elements. 

The matrix elements 
of the NN-like interactions $X_{31}$ and $T^2_\pi (r_{31})$ read 
in momentum space
\begin{eqnarray}
  \label{eq:fourx31}
& &   <(31)2,4 | X_{31} | (31)2,4' > \nonumber \\[5pt]
& & \quad = { \delta(q_4 - q_4 ' ) \over q_4 q_4' } 
          \ { \delta(p_2 - p_2 ' ) \over p_2 p_2' } \
             \delta_{I_4 I_4'} \delta_{I_2 I_2'} \delta_{l_4 l_4'} \delta_{l_2 l_2'} 
             \delta_{J J'} \delta_{M M'} \delta_{j_{31} j_{31}'} \cr
& & \qquad \left[ \tilde Y _{l_{31}} (p_{31},p_{31}') 
                  \ \delta_{l_{31}l_{31}'} \ \delta_{s_{31}s_{31}'} \ (-3+4s_{31})      
               + \tilde T _{l_{31}l_{31}'} (p_{31},p_{31}')  
                    \ \delta_{s_{31}s_{31}'} \ \delta_{s_{31}1} \ S_{l_{31}l_{31}'j_{31}} 
                          \right]    
\end{eqnarray}
and 
\begin{eqnarray}
  \label{eq:shorttq}
& & <(31)2,4 | T_\pi^2(r_{31}) | (31)2,4' >   \nonumber \\[5pt]
& & = { \delta(q_4 - q_4 ' ) \over q_4 q_4' } 
          \ { \delta(p_2 - p_2 ' ) \over p_2 p_2' } \
             \delta_{I_4 I_4'} \delta_{I_2 I_2'} \delta_{l_4 l_4'} \delta_{l_2 l_2'} 
             \delta_{J J'} \delta_{M M'} \delta_{j_{31} j_{31}'} \delta_{l_{31} l_{31}'}
             \delta_{s_{31} s_{31}'} \ \bar T_{l_{31}} (p_{31},p_{31}')   .
\end{eqnarray}
Here the tensor operator can be expressed in simple rational functions of the quantum numbers
\begin{equation}
  \label{eq:tensoroppart}
  \raisebox{3ex}{$S_{l_{31}l_{31}'j_{31}}$ = } 
  \begin{array}{lc}
    \begin{array}{l}
      l_{31}=j_{31}-1 \cr
      l_{31}=j_{31} \cr
      l_{31}=j_{31}+1 \cr
    \end{array}
&   \left[ 
    \begin{array}{ccc}
          - 2 \ { j_{31}-1 \over 2 j_{31}+1 } 
                &  0   & 6 \ { \sqrt{ j_{31}(j_{31}+1)} \over 2j_{31}+1 }  \cr
           0    &  2   &   0  \cr
           6 \ { \sqrt{ j_{31}(j_{31}+1)} \over 2j_{31}+1 }     &  0   
                 &  - 2 \ { j_{31}+2 \over 2 j_{31}+1 }  \cr
    \end{array} \right] \cr
 &  \cr
  &
    \begin{array}{ccc}
      l_{31}'=j_{31}-1 &
      l_{31}'=j_{31}   &
      l_{31}'=j_{31}+1 \cr
    \end{array}
  \end{array}
\end{equation}
We numerically perform the Fourier transformations  
\begin{eqnarray}
  \label{eq:fouryt}
  \tilde Y _{l_{31}} (p_{31},p_{31}') & = & { 2 \over \pi } 
     \ \int_{0}^{\infty} dr \ r^2 \ {\rm j}_{l_{31}} (p_{31}r) \ Y_\pi (r) 
                   \ {\rm j}_{l_{31}} (p_{31}'r) \cr
  \tilde T_{l_{31}l_{31}'} (p_{31},p_{31}') & = & (-) ^{l_{31}-l_{31}'} 
     \ { 2 \over \pi } 
     \ \int_{0}^{\infty} dr \ r^2 \ {\rm j}_{l_{31}} (p_{31}r) \ T_\pi (r) 
                   \ {\rm j}_{l_{31}'} (p_{31}'r)  \cr 
  \bar T_{l_{31}} (p_{31},p_{31}') & = &  
     { 2 \over \pi } 
     \ \int_{0}^{\infty} dr \ r^2 \ {\rm j}_{l_{31}} (p_{31}r) \ T^2_\pi (r) 
                   \ {\rm j}_{l_{31}} (p_{31}'r)   
\end{eqnarray}
with the usual  spherical Bessel functions ${\rm j}_{l}(x)$.  

Because these NN-like interactions are all symmetric with respect 
to an interchange of the subsystem particles, the matrix elements 
for the $X_{23}$ and $T^2(r_{23})$ equal those for $X_{31}$ and $T^2(r_{31})$,
respectively
 
\bibliography{literatur}

\end{document}